\documentclass[letterpaper,american,english]{emulateapj}
\usepackage[T1]{fontenc}
\usepackage[latin9]{inputenc}
\usepackage{amssymb}
\usepackage{graphicx}

\makeatletter

\providecommand{\tabularnewline}{\\}

\newcommand{\source}{IGR~J17062-6143}
\shorttitle{Reflection and a Burst from IGR~J17062-6143}
\shortauthors{Keek, Iwakiri, Serino, et al.}

\makeatother

\usepackage{babel}
\begin{document}

\title{X-ray Reflection and an Exceptionally Long Thermonuclear Helium Burst
from \source{}}

\author{L.~Keek,\altaffilmark{1,2} W.~\foreignlanguage{american}{Iwakiri},\altaffilmark{3}
M.~Serino,\altaffilmark{3} D.\,R.~Ballantyne,\altaffilmark{4}
J.\,J.\,M.~in~'t~Zand,\altaffilmark{5} T.\,E.~Strohmayer\altaffilmark{1}}

\altaffiltext{1}{X-ray Astrophysics Laboratory, Astrophysics Science Division, NASA/GSFC, Greenbelt, MD 20771} 
\altaffiltext{2}{CRESST and the Department of Astronomy, University of Maryland, College Park, MD 20742}
\altaffiltext{3}{MAXI team, RIKEN, 2-1 Hirosawa, Wako, Saitama 351-0198, Japan}
\altaffiltext{4}{Center for Relativistic Astrophysics, School of Physics, Georgia Institute of Technology, 837 State Street, Atlanta, GA 30332-0430}
\altaffiltext{5}{SRON Netherlands Institute for Space Research, Sorbonnelaan 2, 3584 CA Utrecht, The Netherlands}

\email{laurens.keek@nasa.gov}
\begin{abstract}
Thermonuclear X-ray bursts from accreting neutron stars power brief
but strong irradiation of their surroundings, providing a unique way
to study accretion physics. We analyze \emph{MAXI}/GSC and \emph{Swift}/XRT
spectra of a day-long flash observed from \source{} in 2015. It is
a rare case of recurring bursts at a low accretion luminosity of $0.15\%$
Eddington. Spectra from \emph{MAXI}, \emph{Chandra}, and \emph{NuSTAR}
observations taken between the 2015 burst and the previous one in
2012 are used to determine the accretion column. We find it to be
consistent with the burst ignition column of $5\times10^{10}\,\mathrm{g\,cm^{-2}}$,
which indicates that it is likely powered by burning in a deep helium
layer. The burst flux is observed for over a day, and decays as a
straight power law: $F\propto t^{-1.15}$. The burst and persistent
spectra are well described by thermal emission from the neutron star,
Comptonization of this emission in a hot optically thin medium surrounding
the star, and reflection off the photoionized accretion disk. At the
burst peak, the Comptonized component disappears, when the burst may
dissipate the Comptonizing gas, and it returns in the burst tail.
The reflection signal suggests that the inner disk is truncated at
$\sim10^{2}$ gravitational radii before the burst, but may move closer
to the star during the burst. At the end of the burst, the flux drops
below the burst cooling trend for $2$ days, before returning to the
pre-burst level.
\end{abstract}


\section{Introduction}

Mass transfer from a binary companion star onto a neutron star can
enrich the latter's surface in hydrogen and/or helium. The strong
compression by the neutron star's gravity induces nuclear fusion in
the surface layer. If the nuclear burning proceeds as a thermonuclear
runaway, the accreted material burns within seconds, powering a bright
Type I X-ray burst \citep{Grindlay1976,1976Belian,Woosley1976,Maraschi1977}.
Most of the thousands of observed bursts last $\sim10-100\,\mathrm{s}$
\citep[e.g.,][]{Cornelisse2003,Galloway2008catalog}. Intermediate
duration bursts ($\sim100-1000\,\mathrm{s}$) and superbursts ($\gtrsim1000\,\mathrm{s}$)
are observed relatively rarely. They are thought to be powered by
the unstable burning of deep layers of helium and carbon, respectively
(\citealp[e.g.,][]{Keek2008int..work}; for ignition conditions and
energetics see, \citealp[e.g.,][]{Cumming2006}). Several tens of
bursts have been detected from both categories, with durations of
minutes to hours that allow for higher quality spectra to be collected.
For example, reflection features were detected for two superbursts
\citep{Strohmayer2002,Ballantyne2004,Keek2014sb2} and an intermediate
duration burst \citep{Degenaar2013}, and superexpansion as well as
strong flux variability on timescales of seconds were present in several
intermediate duration bursts \citep{Molkov2005,Zand2011,Degenaar2013}
and one superburst \citep{Strohmayer2002,Zand2010}. These are instances
where X-ray bursts have a strong impact on the accretion environment
around the neutron star, and demonstrate how X-ray bursts can be employed
to study accretion physics \citep[e.g.,][]{Ballantyne2005}.

The majority of bursts has been observed with instruments sensitive
above $\sim2\,\mathrm{keV}$ and with modest spectral resolution.
The burst signal in the soft X-ray band is relatively unexplored,
even though, for instance, the burst reflection signal may dominate
this band \citep{Ballantyne2004models}. A bright burst observed from
\object{\source{}} in 2012 with the \emph{Swift} observatory exhibited
spectral emission features below $2\,\mathrm{keV}$ \citep{Degenaar2013}.
On 11/3/2015 the nova alert system \citep{Negoro2016} of the \emph{Monitor
of All-sky X-ray Image} (MAXI) detected a second powerful burst from
this source \citep{Negoro2015}. A campaign of follow-up observations
was subsequently performed by \emph{Swift} \citep{Iwakiri2015}. In
this paper we analyze the observations of the 2015 burst to investigate
its impact on its surroundings.

\source{} was discovered in 2006 at the start of an outburst \citep{Churazov2007,Ricci2008ATel,Remillard2008ATel}.
Since then it has continued to be active at a low flux. \emph{Chandra}
gratings spectra of the persistent emission exhibit emission and absorption
lines \citep{Degenaar2016}. \emph{NuSTAR }observations revealed reflection
features, that indicate that the accretion disk is highly ionized
and truncated far from the neutron star surface at $\gtrsim10^{2}\,R_{\mathrm{g}}$
($R_{\mathrm{g}}=GM/c^{2}$ is the gravitational radius). \citet{Degenaar2016}
discuss how a strong magnetic field could truncate the disk, form
a radiatively inefficient accretion flow, and act as a propeller to
drive an outflow \citep[for a review see][]{DAngelo2015}. However,
our knowledge of \source{} remains limited. The neutron star spin,
the composition and inclination of the disk, and the binary period
are unknown.

After describing the employed observations and spectral models (Section~\ref{sec:Observations-and-Spectral}),
we jointly analyze the persistent \emph{MAXI}, \emph{Chandra}, and
\emph{NuSTAR} spectra to determine the time-averaged persistent flux
between the two bursts (Section~\ref{sec:pers_spectra}). The \emph{MAXI}
and \emph{Swift} burst spectra are analyzed to establish the properties
of the thermonuclear flash (Section~\ref{sec:Analysis-of-Burst}).
In both cases we fit a simple phenomenological model as well as a
more physically motivated model that includes Comptonization and photoionized
reflection. The results show that this burst had a strong influence
on the accretion geometry (Section~\ref{sec:Discussion}), and we
conclude that it is among the most powerful helium flashes observed
from accreting neutron stars (Section~\ref{sec:Conclusions-and-Outlook}).

\section{Observations and Spectral Models}

\label{sec:Observations-and-Spectral}

\subsection{Burst Observations}

\begin{table}
\caption{\label{tab:Observations}Burst Observations}

\begin{centering}
\begin{tabular}{cccr@{\extracolsep{0pt}.}l}
\hline 
Instrument & ObsID & Mode & \multicolumn{2}{c}{Exposure (ks)}\tabularnewline
\hline 
\emph{MAXI}/GSC & 11/3/2015 10:29 UT &  & \multicolumn{2}{c}{$0.06$}\tabularnewline
 & 11/3/2015 12:03 UT &  & \multicolumn{2}{c}{$0.06$}\tabularnewline
\hline 
\emph{Swift}/XRT$^{a}$ & 00037808006 & PC & \multicolumn{2}{c}{$1.7$}\tabularnewline
 & 00037808008 & WT & \multicolumn{2}{c}{$3.4$}\tabularnewline
 & 00037808008 & PC & \multicolumn{2}{c}{$1.8$}\tabularnewline
 & 00037808009 & PC & \multicolumn{2}{c}{$5.6$}\tabularnewline
 & 00037808010 & WT & \multicolumn{2}{c}{$4.0$}\tabularnewline
 & 00037808012 & WT & \multicolumn{2}{c}{$0.9$}\tabularnewline
 & 00037808015 & WT & \multicolumn{2}{c}{$6.2$}\tabularnewline
 & 00037808016 & WT & \multicolumn{2}{c}{$6.7$}\tabularnewline
 & 00037808017 & WT & \multicolumn{2}{c}{$6.9$}\tabularnewline
 & 00037808018 & WT & \multicolumn{2}{c}{$6.7$}\tabularnewline
 & 00037808019 & WT & \multicolumn{2}{c}{$7.2$}\tabularnewline
 & 00037808020 & WT & \multicolumn{2}{c}{$4.3$}\tabularnewline
 & 00037808021 & PC & \multicolumn{2}{c}{$4.5$}\tabularnewline
 & 00037808022 & PC & \multicolumn{2}{c}{$5.0$}\tabularnewline
 & 00037808023 & PC & \multicolumn{2}{c}{$3.9$}\tabularnewline
 & 00037808024 & PC & \multicolumn{2}{c}{$6.5$}\tabularnewline
\hline 
\end{tabular}
\par\end{centering}

$^{a}$ \emph{Swift}/XRT observations started on 11/3/2015 13:18 UT
and ended at 11/15/2015 06:07 UT. ObsIDs 00037808006 to 00037808012
cover the burst decay (up to $2\times10^{5}\,\mathrm{s}$ in Figure~\ref{fig:lcv}).
\end{table}

\emph{MAXI} \citep{Matsuoka2009MAXI} was installed on the International
Space Station (ISS) in 2009. We employ data from the Gas Slit Camera
\citep[GSC;][]{Mihara2011maxiGSC,Sugizaki2011maxiGSC}, which consists
of $12$ xenon-filled proportional counters that are sensitive in
the $2-30\,\mathrm{keV}$ energy range and have a combined collecting
area of $5350\,\mathrm{cm^{2}}$. A slit and slat collimator restricts
the field of view to a narrow elongated region of $3^{\circ}\times80^{\circ}$.
$85\%$ of the sky is scanned each $92$ minute orbit of the ISS.
At the time of the burst, \source{} dominated the X-ray flux near
the source position, and no other nearby transients were active. We
extract source spectra for the triggering GSC scan on 11/3/2015 10:29
UT as well as the subsequent scan (Table~\ref{tab:Observations}).\footnote{MAXI spectra are publicly available at \url{http://maxi.riken.jp/top/}.}
A scan preceding the trigger by $92$ minutes had not detected the
burst. The source is visible for one minute during each scan, and
the effective area peaks at $3\,\mathrm{cm^{2}}$ in the middle of
the scans. The first spectrum has $893$ counts, whereas the second
consists of only $90$ counts. The background and instrument response
are modeled with tools provided by the instrument team \citep{Sugizaki2011maxiGSC}.

\begin{figure*}
\includegraphics{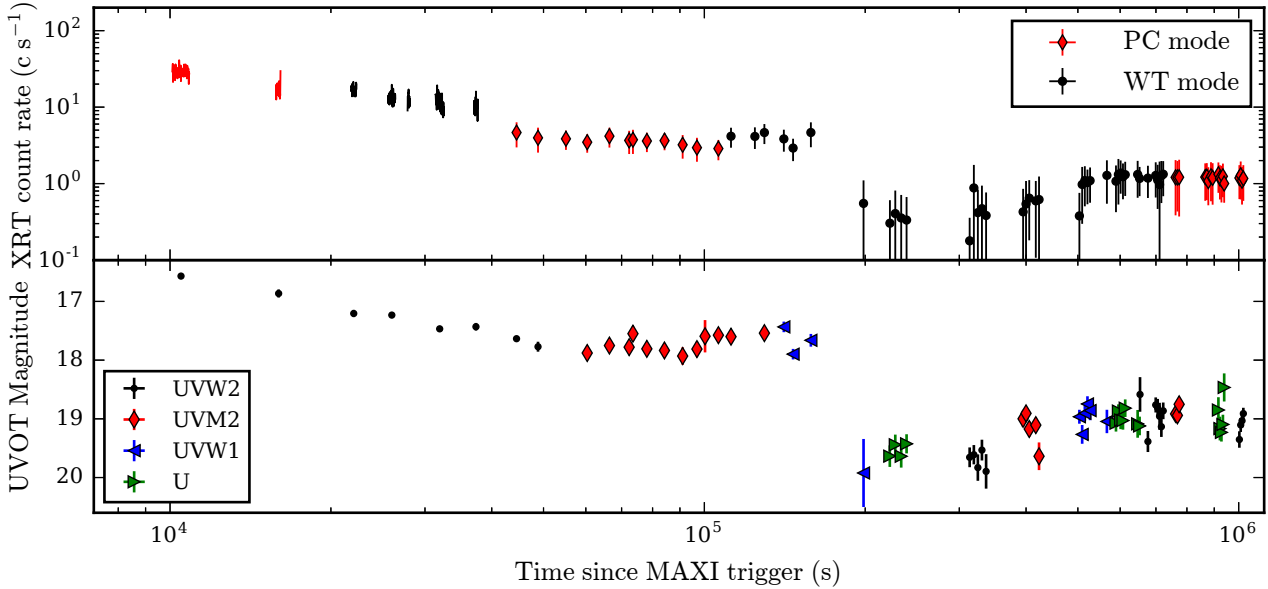}

\caption{\label{fig:lcv}(Top) Light curve of the XRT count rates as a function
of time since the \emph{MAXI} burst detection. For the first $11\,\mathrm{hr}$
we show the data at $10\,\mathrm{s}$ time resolution, whereas for
later observations each data point is the average per orbit and error
bars indicate the root-mean-squared of the data at $10\,\mathrm{s}$
resolution. The first 2 orbits were observed in PC mode, the next
5 in WT mode, and for subsequent orbits the mode is indicated by symbols.
The count rates in PC mode are corrected for the loss of effective
area due to pile up with the \textsc{xrtlccorr} tool. (Bottom) Magnitude
of UVOT detections in the indicated filters.}
\end{figure*}
 The \emph{Swift} observatory was launched in 2004 \citep{Gehrels2004}.
Its main pointed X-ray instrument is the X-Ray Telescope \citep[XRT;][]{Burrows2005xrt}.
The XRT is a CCD imager sensitive in the $0.2-10\,\mathrm{keV}$ band
with an effective area of $120\,\mathrm{cm^{2}}$ at $1.5\,\mathrm{keV}$.
Starting $3$ hours after the first \emph{MAXI}/GSC scan, \emph{Swift}
performed a series of pointed observations of \source{} \citep{Iwakiri2015}.
A total of $75.3\,\mathrm{ks}$ were collected over $11.7$ days (Table~\ref{tab:Observations};
Figure~\ref{fig:lcv}). The XRT observations were performed in either
Photon Counting (PC) mode or Windowed Timing (WT) mode. Whereas in
PC mode the full CCD image is stored, in WT mode a reduced 1D image
is read out to increase the time resolution. We use \textsc{xrtpipeline}
to extract spectra in the $0.5-10\,\mathrm{keV}$ band and light curves
from a circular region with a radius of $70.8$ arcsec ($30$ pixels)
centered on the source, and from an off-source region of the same
size as background. The standard selection of event grades are used:
$0-12$ for PC mode and $0-2$ for WT mode. The PC mode data suffer
from pile-up. We exclude the piled-up center of the point-spread-function.
The size of this region is determined by comparing the observed spatial
distribution of events to a King profile that describes the expected
point-spread-function \citep{Moretti2005}: the excluded regions have
a radius ranging from $20$ arcsec at the highest flux to $9$ arcsec
at the lowest flux. The detector light curves are inspected for background
flares. The ancillary response is generated by \textsc{xrtpipeline},
and we employ the appropriate response matrices provided by the instrument
team: swxpc0to12s6\_20130101v014.rmf for PC mode and swxwt0to2s6\_20131212v015.rmf
for WT mode. The data are labeled with an Observation Identifier (ObsID),
and each ObsID represents several consecutive satellite orbits. For
the first two ObsIDs, which cover the initial burst decay, we create
separate spectra for each orbit. For the rest, one spectrum per ObsID
is generated. The resulting XRT spectra each have between $840$ and
$13,175$ counts, with a median value of $5003$ counts. In Section~\ref{sub:Burst-properties}
we demonstrate that these spectra provide sufficient time resolution
to resolve the cooling trend of the burst. 

\emph{Swift}'s other pointed instrument is the Ultraviolet/Optical
Telescope \citep[UVOT;][]{Roming2005uvot}, which records CCD images
in the $170-650\,\mathrm{nm}$ wavelength range. UVOT has seven filters
to select a narrow wavelength interval from this range, four of which
were used in the different pointings (Figure~\ref{fig:lcv} bottom).
UVOT magnitudes are extracted using the \textsc{uvotmaghist} tool
from the same source and background locations as employed for the
XRT. A standard aperture size of $5$~arcsec is used, and approximate
aperture corrections are applied by the \textsc{uvotsource} tool.

\subsection{Observations of Persistent Emission}

\begin{figure}
\includegraphics{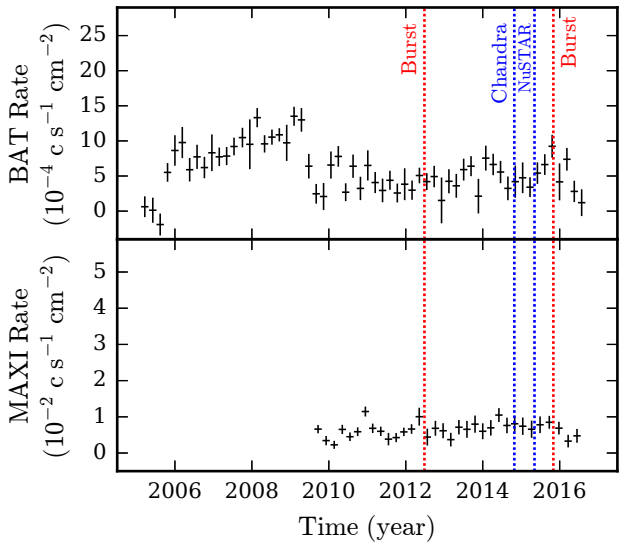} 

\caption{\label{fig:persistent_lcv} Light curve of the persistent emission
from \source{} since its discovery in 2006. The panels show the photon
flux as observed with \emph{Swift}/BAT (top; $15-150\,\mathrm{keV}$
band pass) and \emph{MAXI}/GSC (bottom; $2-10\,\mathrm{keV}$ band
pass) at a $70$~day time resolution. Dotted lines indicate the times
of the \emph{Chandra} and \emph{NuSTAR} observations as well as the
two burst observations, where the last burst is studied in this paper.
In Section~\ref{sec:pers_spectra} we study the combined \emph{MAXI}
spectrum accumulated between the two bursts.}
\end{figure}

\begin{table}
\caption{\label{tab:persistent-Observations}Observations of Persistent Emission}

\centering{}%
\begin{tabular}{ccr@{\extracolsep{0pt}.}l}
\hline 
Instrument & ObsID & \multicolumn{2}{c}{Exposure (ks)}\tabularnewline
\hline 
\emph{MAXI}/GSC & 6/26/2012 - 11/2/2015 & \multicolumn{2}{c}{$2,302.5$}\tabularnewline
\hline 
\emph{Chandra}/HETGS & 15749 (10/25/2014) & \multicolumn{2}{c}{$29.3$}\tabularnewline
 & 17543 (10/27/2014) & \multicolumn{2}{c}{$63.5$}\tabularnewline
\hline 
\emph{NuSTAR} & 30101034002 (05/06/2015) & \multicolumn{2}{c}{$70.1$}\tabularnewline
\hline 
\end{tabular}
\end{table}
 \emph{Swift}'s Burst Alert Telescope \citep[BAT;][]{Barthelmy2005}
is a coded-mask imager with a wide field of view of $1.4\,\mathrm{sr}$
and a $15-150\,\mathrm{keV}$ band pass. We employ BAT in combination
with \emph{MAXI} to illustrate the long term evolution of the persistent
flux (Figure~\ref{fig:persistent_lcv}).\footnote{\emph{Swift}/BAT was not operational at the time of the 2015 burst
(see GCN circulars 18562 and 18610)} The source was first detected at the onset of its outburst in 2006,
and it has been accreting continuously at a low rate. \emph{MAXI}
operations started in 2009. We are especially interested in the period
between the burst observed in 2012 \citep{Degenaar2013} and the one
in 2015 (studied in this paper). To quantify the variability in that
time interval, we take the root mean squared of the photon flux at
a $70$~day resolution: it is $38\%$ of the mean for BAT and $22\%$
for\emph{ MAXI}. The corresponding hardness ratio from the $4-20\,\mathrm{keV}$
and $2-4\,\mathrm{keV}$ \emph{MAXI} bands does not exhibit significant
variability. We take this as an indication that the spectral shape
did not change substantially, and we extract one \emph{MAXI}/GSC spectrum
from all observations in that interval combined (Table~\ref{tab:persistent-Observations}).
We restrict the energy range to $2-10\,\mathrm{keV}$ where the source
is detected most clearly, collecting a total of $5.5\times10^{3}$
net source counts. In the same period, the source was also observed
with the following instruments.

The \emph{Chandra X-ray Observatory} \citep{Weisskopf2000} was launched
in 1999. \emph{Chandra} observed the source on 10/25/2014 and 10/27/2014
for a total of $93\,\mathrm{ks}$ with the High Energy Transmission
Grating Spectrometer \citep[HETGS;][]{Canizares2005}, which includes
the High Energy Grating (HEG) and the Medium Energy Grating (MEG).
We use the spectra and response matrices provided in the Chandra Grating-Data
Archive and Catalog \citep[TGCat;][]{Huenemoerder2011}.\footnote{The Chandra data products and further details on their extraction
are available at \url{http://tgcat.mit.edu}.} The data products were extracted using a narrow mask for better flux
correction of the HEG below $6.9\,\mathrm{keV}$. The background was
extracted from an off-source position. No significant variability
is apparent during the pointings. We will analyze the $0.5-10\,\mathrm{keV}$
spectra of the $+1,-1$ orders of the MEG and HEG for both pointings,
which have a total of $1.8\times10^{5}$ counts ($91\%$ of the counts
in all orders). The MEG and HEG spectra of both orders for each pointing
will be fit jointly.

The \emph{Nuclear Spectroscopic Telescope Array} \citep[NuSTAR;][]{Harrison2013}
observes the hard X-ray sky since 2012 with its focusing optics and
two imaging Focal-Plane Modules: FPMA and FPMB. \emph{NuSTAR} observed
\source{} on 05/06/2015 for $70.1\,\mathrm{ks}$. The data is reprocessed
with \textsc{nupipeline} version 0.4.5 using calibration data with
the time stamp 07/31/2016, creating spectra for both modules in the
$3-50\,\mathrm{keV}$ band from a circular extraction region with
a radius of $65$~arcsec. The combined FPMA and FPMB spectra have
$1.5\times10^{5}$ counts. Background spectra are extracted from an
off-source position, and we find that the background dominates the
signal at energies in excess of $\gtrsim50\,\mathrm{keV}$. During
the pointing, the count rate is consistent with being constant.

\subsection{Interstellar Absorption\label{sub:Interstellar-Absorption}}

As part of our spectral model, interstellar absorption is described
by the Tübingen-Boulder model with abundances from \citet{Wilms2000}.
The tool \textsc{NHtot}\,\footnote{See \url{http://www.swift.ac.uk/analysis/nhtot/}}
calculates the Galactic absorption column of atomic \citep{Kalberla2005}
and molecular \citep{Schlegel1998} hydrogen using radio and infrared
maps, respectively \citep{Willingale2013}. In the direction of \source{},
within $1^{\circ}$ from the source, we find a mean value of $N_{\mathrm{H}}=1.58\times10^{21}\,\mathrm{cm^{-2}}$. 

For bright low-mass X-ray binaries (LMXBs), absorption lines and edges
have been used to quantify the interstellar absorption column \citep[e.g.,][]{Pinto2010}.
A study of the Chandra gratings spectra of the persistent emission
did not find significant absorption features that could be used for
this purpose \citep{Degenaar2016}. Alternatively, $N_{\mathrm{H}}$
may be determined as part of the fit to the continuum spectrum. Indeed,
$N_{\mathrm{H}}$ measured from XRT spectra of the 2012 burst from
\source{} is consistent with \textsc{NHtot} \citep{Degenaar2013}.
The best-fitting value is, however, dependent on the model for the
continuum spectrum, because a model that adds more flux at low energies
($E\lesssim2\,\mathrm{keV}$) requires a larger $N_{\mathrm{H}}$
to compensate \citep[see, e.g., the fits by][]{Degenaar2016}. Because
$N_{\mathrm{H}}$ cannot be robustly constrained by the X-ray spectra,
we fix $N_{\mathrm{H}}$ to the value from \textsc{NHtot}, under the
assumption that there is no substantial contribution from a local
absorber \citep[e.g.,][]{Degenaar2016}.

\subsection{Models of Photoionized Reflection\label{sub:Models-of-Photoionized}}

\begin{figure}
\includegraphics{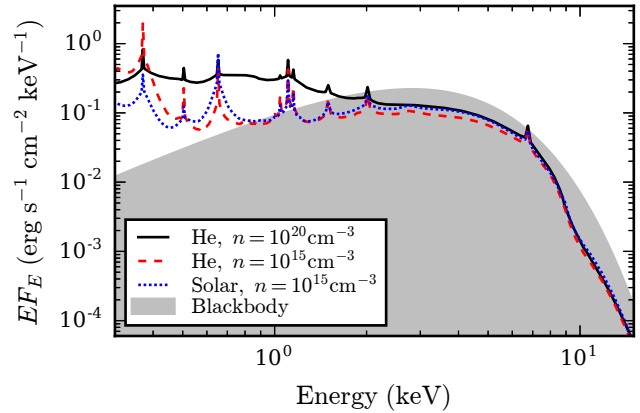}

\caption{\label{fig:reflection_models}Comparison of the photon flux, $F_{E}$,
multiplied by the photon energy as a function of energy as predicted
by a blackbody with temperature $kT=1.0\,\mathrm{keV}$ (top of the
shaded region) and three reflection models. The reflection models
reprocess the blackbody for a disk with $\log\xi=3.0$, either solar
composition or helium with solar metal abundance, and given number
density $n$. All models have been normalized such that the integrated
$EF_{E}$ is unity.}
\end{figure}
We employ models of reflection off a photoionized accretion disk that
is illuminated by either a power law or a blackbody spectrum. Both
models implement the disk as a slab of constant number density $n$
and with ionization parameter $\xi\equiv4\pi F/n$, where $F$ is
the irradiating flux. We will report $\log\xi$, with $\xi$ in units
of $\mathrm{erg\,s^{-1}\,cm}$.

The reflection spectrum has a dependence on the composition of the
reflecting material. For the ultra-compact X-ray binary (UCXB) 4U~1820\textendash 30,
\citet{Ballantyne2004models} calculated models of a reflected blackbody
assuming a helium-rich composition with metals at solar abundance.
The accretion composition of \source{} is unknown at present, but
intermediate duration bursts and accretion at a constant low rate
have been associated with UCXBs \citep[e.g.,][]{intZand2007}. It
is, therefore, plausible that \source{} is of a similar nature as
4U~1820\textendash 30, and the same reflection models are also applicable
here. The models assume a density of $n=10^{15}\,\mathrm{cm^{-3}}$,
although $n=10^{20}\,\mathrm{cm^{-3}}$ may be more realistic for
the inner disk. The density dependence is most pronounced for $E\lesssim3\,\mathrm{keV}$
(Figure~\ref{fig:reflection_models}; \citealt{Ballantyne2004models}).
It is unimportant when fitting, e.g., \emph{RXTE}/PCA data \citep{Ballantyne2004,Keek2014sb2}
or \emph{MAXI} data, and we use the \citet{Ballantyne2004models}
models for the latter. For the XRT spectra, however, we calculate
a new grid of models with $n=10^{20}\,\mathrm{cm^{-3}}$, using the
same procedure as \citet{Ballantyne2004models}. The new grid covers
a range of blackbody temperatures $0.2\,\mathrm{keV}\le kT\le1.2\,\mathrm{keV}$
as well as $1.5\le\log\xi\le3.0$.

For reflection of a power law we employ version 0.4c of the \texttt{relxill}
model \citep{Garcia2014relxill,Dauser2014}. \texttt{relxill} provides
the illuminating power law with photon index $\Gamma$ and a high
energy cutoff as well as the \texttt{xillver} model of reflection
off a photoionized accretion disk \citep{Garcia2013}. The flux ratio
of the reflection and illumination components is given by the reflection
fraction, $f_{\mathrm{refl}}$. \texttt{xillver} assumes a density
of $n=10^{15}\,\mathrm{cm^{-3}}$ and a composition based on solar
with a variable iron abundance. Unfortunately, the composition and
density do not match the values that we preferred for blackbody reflection.
Therefore, we only apply this model to $E>3\,\mathrm{keV}$, where
the effect of these parameters is minimal (see Figure~\ref{fig:reflection_models}
for an illustration using blackbody reflection).

\texttt{relxill} further takes into account relativistic effects that
smooth the reflection spectrum using the \texttt{relline} code \citep{Dauser2010relline},
depending on the inclination angle of the disk with respect to the
line of sight and the emissivity profile of the disk. We apply the
same smoothing to the blackbody reflection models using the \texttt{relconv}
convolution model, which is also based on the \texttt{relline} code.

\section{Analysis of Persistent Emission}

\label{sec:pers_spectra}

We investigate the persistent emission between the 2012 and 2015 bursts
to measure the time-averaged persistent flux and to quantify the spectral
shape for comparison with the burst spectra in Section~\ref{sec:Analysis-of-Burst}.
The \emph{MAXI} data collected throughout this period (Table~\ref{tab:persistent-Observations})
provide a good measure of the time-averaged flux level, but its modest
spectral quality does not strongly constrain the spectral shape (Figure~\ref{fig:The-persistent-spectrum}
top). Because the flux exhibits only minor variability in that period
(Figure~\ref{fig:persistent_lcv}), we assume that the spectral shape
was unchanged, and we employ the \emph{Chandra}, and \emph{NuSTAR}
spectra listed in Table~\ref{tab:persistent-Observations} to study
the shape of the spectral components. The \emph{Chandra}, and \emph{NuSTAR}
spectra are part of previous study by \citet{Degenaar2016}, who describe
the spectrum with a blackbody, a power law, and photoionized reflection
of the power law. We employ the same model (Section~\ref{sub:Phenomenological-fit},
\ref{sub:Relxill-Reflection-Model}), but make different assumptions
for certain parameters, most importantly the interstellar absorption
column $N_{\mathrm{H}}$. \citet{Degenaar2016} fit for $N_{\mathrm{H}}$,
obtaining a relatively large error, and this influences the absorption
correction of the flux as well as the parameter values of the continuum
components. We avoid this by fixing $N_{\mathrm{H}}$ (Section~\ref{sub:Interstellar-Absorption}).
Furthermore, we improve the bolometric correction of the persistent
flux by modifying the spectral model to include a Comptonization component
(Section~\ref{sub:Simpl-Comptonization-Model}). We will use the
same assumptions when analyzing the burst spectra (Section~\ref{sec:Analysis-of-Burst}),
such that we have a consistent picture of both the persistent and
burst emission.

The persistent spectra are analyzed with \textsc{XSPEC} v.12.9.0i
\citep{Arnaud1996}. For all spectra, neighboring spectral bins that
have fewer than $15$ counts are grouped to ensure that the uncertainties
in the data points are close to Gaussian. We use $\chi^{2}$ statistics
and report the $1\,\sigma$ uncertainties in the fit parameters.

\subsection{Phenomenological fit\label{sub:Phenomenological-fit}}

All persistent \emph{MAXI}, \emph{Chandra}, and \emph{NuSTAR} spectra
are fit jointly, allowing for multiplication factors between \emph{MAXI},
\emph{NuSTAR} FPMA, \emph{NuSTAR} FPMB, \emph{Chandra} observation
15749, and \emph{Chandra} observation 17543 (using \textsc{XSPEC}
model \texttt{constant}). The factor for the latter is fixed to unity.
We first fit a model that includes a blackbody with temperature $kT$
and normalization $K_{\mathrm{bb}}$, as well as a power law with
photon index $\Gamma$ . Instead of the power law normalization, we
report the $0.5-10\,\mathrm{keV}$ unabsorbed flux of the power law
provided by a \texttt{cflux} component in \textsc{XSPEC}. Interstellar
absorption is implemented as described in Section~\ref{sub:Interstellar-Absorption}.
The complete \textsc{XSPEC} model is \texttt{constant{*}TBabs(bbodyrad
+ cflux{*}powerlaw)}. 
\begin{figure}
\includegraphics{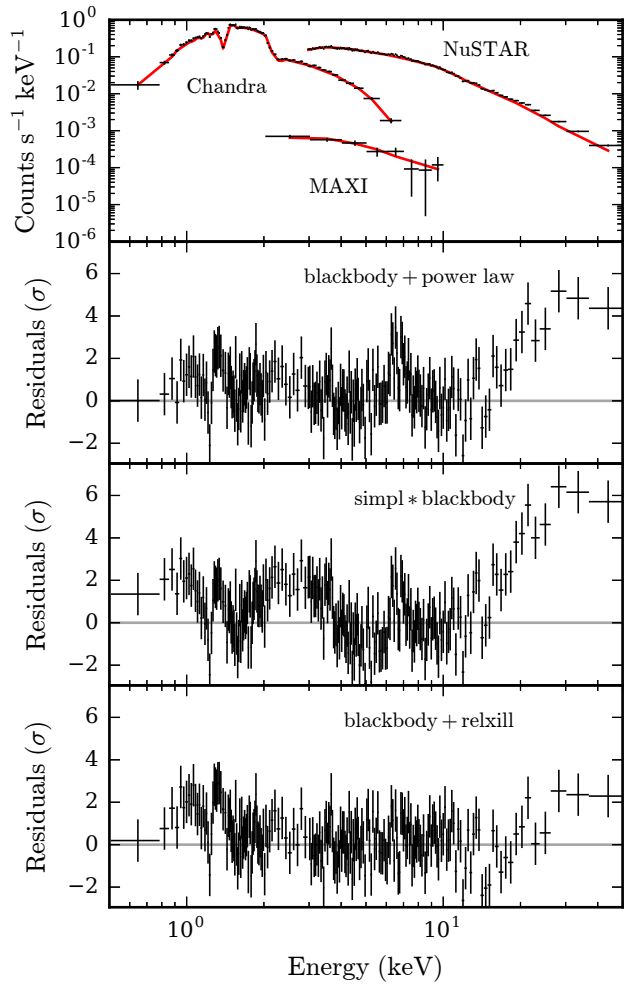}

\caption{\label{fig:The-persistent-spectrum}The persistent spectrum (top panel)
and residuals of fits with the indicated models (other panels). We
show the spectra of MEG order 1 from \emph{Chandra} pointing 17543,
\emph{NuSTAR} FPMA, and \emph{MAXI}/GSC, and the spectra have been
rebinned for clarity. The residuals include the MEG up to $3\,\mathrm{keV}$
and the FPMA in its full $3-50\,\mathrm{keV}$ band. The solid line
in the top panel indicates the fit with the blackbody+power law model,
whose residuals in the second panel exhibit an Fe K$\alpha$ line
near $6.4\,\mathrm{keV}$ and a Compton hump above $20\,\mathrm{keV}$.
Those features are largely described by the relxill component of photoionized
reflection (bottom panel).}
\end{figure}
It provides a reasonable fit to the data (Figure~\ref{fig:The-persistent-spectrum}
top). The residuals of the \emph{Chandra}/MEG data show some features
near $1\,\mathrm{keV}$, but the HEG is less sensitive at those energies
and its spectra do not exhibit them. The residuals for the \emph{NuSTAR}
data show an emission line near $6.4\,\mathrm{keV}$ and a broad excess
at $E>20\,\mathrm{keV}$, which may be the Fe~K$\alpha$ line complex
and the Compton hump, respectively (second panel of Figure~\ref{fig:The-persistent-spectrum}),
that are known to be produced by photoionized reflection. For \emph{NuSTAR}
we exclude the parts of the spectra where reflection features appear:
$6.0\,\mathrm{keV}<E<7.5\,\mathrm{keV}$ and $20.0\,\mathrm{keV}<E<50\,\mathrm{keV}$.
In principle the reprocessing of the spectrum by reflection can also
influence the measured continuum parameters such as $\Gamma$, but
we find this not to be an important effect in this case \citep[see also the discussion in][]{Keek2016agnrefl}.
\begin{table}
\caption{\label{tab:Fit-all-po}Fit of the Persistent Spectra with a Blackbody
and Power Law}

\begin{centering}
\begin{tabular}{cc}
\hline 
\texttt{bbodyrad} & \tabularnewline
$kT\,(\mathrm{keV)}$ & $0.456\pm0.005$\tabularnewline
$K_{\mathrm{bb}}\,(\mathrm{km/10\,kpc})^{2}$ & $48\pm3$\tabularnewline
\hline 
\texttt{powerlaw} & \tabularnewline
$\Gamma$ & $2.159\pm0.008$\tabularnewline
$F_{\mathrm{po}}\,(\mathrm{10^{-10}erg\,s^{-1}\,cm^{-2}})$$^{a}$ & $1.03\pm0.02$\tabularnewline
\hline 
\texttt{constant} & scaling relative to \emph{Chandra} 17543\tabularnewline
\emph{Chandra} 15749 & $0.974\pm0.007$\tabularnewline
\emph{NuSTAR}/FPMA & $0.923\pm0.010$\tabularnewline
\emph{NuSTAR}/FPMB & $0.907\pm0.010$\tabularnewline
\emph{MAXI} & $0.54\pm0.04$\tabularnewline
\hline 
$\chi_{\nu}^{2}\,(\mathrm{degrees\,of\,freedom})$ & $0.58\,(9455)$\tabularnewline
\hline 
\end{tabular}
\par\end{centering}

$^{a}$ Unabsorbed flux in the $0.5-10\,\mathrm{keV}$ band for \emph{Chandra}
ObsID 17543
\end{table}
 The best-fitting parameter values are presented in Table~\ref{tab:Fit-all-po}.
The scaling factor for \emph{Chandra} pointing 15749 is consistent
with the ratio of the count rates in the two Chandra pointings of
$0.981$. The ratio of the factors for \emph{NuSTAR} FPMA and FPMB
is $1.02\pm0.02$: the two spectra are consistent. The scaling factor
for \emph{MAXI} indicates that the long-term average flux was a bit
lower than at the times of the \emph{Chandra} and \emph{NuSTAR} pointings.
Furthermore, the spectrum is dominated by the power law: the ratio
of the $0.5-10\,\mathrm{keV}$ power law flux and the bolometric blackbody
flux is $4.6\pm0.3$.

\subsection{Simpl Comptonization Model\label{sub:Simpl-Comptonization-Model}}

To measure the bolometric flux, the model needs to be extrapolated
outside of the combined instrument bands. The power law component
poses a problem, as it strongly increases toward lower energies. A
power law component from accreting compact objects is often explained
as the result of inverse Compton scattering in an accretion disk corona.
In our case, flux from the blackbody component could be Comptonized
by hot electrons in a corona. In the spectral model we replace the
power law by the \texttt{simpl} model \citep{Steiner2009} convolved
with the blackbody. This is an empirical Comptonization model that
takes a ``scattering fraction'' $f_{\mathrm{sc}}$ of the blackbody
flux, and produces a power law with photon index $\Gamma$ toward
higher energies from Compton upscattering. Toward lower energies,
the downscattered flux falls off quickly with energy. A \texttt{cflux}
component is used to measure the total unabsorbed bolometric flux
in the $0.01-100\,\mathrm{keV}$ range. The complete \textsc{XSPEC}
model becomes \texttt{constant{*}TBabs{*}cflux{*}simpl{*}bbodyrad}.
\begin{table}
\caption{\label{tab:Fit-all-simpl}Fit of the Persistent Spectra with a Blackbody
and Simpl}

\begin{centering}
\begin{tabular}{cc}
\hline 
\texttt{bbodyrad} & \tabularnewline
$kT\,(\mathrm{keV)}$ & $0.298\pm0.006$\tabularnewline
\hline 
\texttt{simpl} & \tabularnewline
$\Gamma$ & $2.250\pm0.007$\tabularnewline
$f_{\mathrm{sc}}$ & $0.664\pm0.013$\tabularnewline
\hline 
\texttt{constant} & scaling relative to \emph{Chandra} 17543\tabularnewline
\emph{Chandra} 15749 & $0.973\pm0.007$\tabularnewline
\emph{NuSTAR} FPMA & $0.882\pm0.010$\tabularnewline
\emph{NuSTAR} FPMB & $0.865\pm0.010$\tabularnewline
\emph{MAXI} & $0.54\pm0.04$\tabularnewline
\hline 
$F_{\mathrm{bol}}\,(\mathrm{10^{-10}erg\,s^{-1}\,cm^{-2}})$ $^{a}$ & $1.70\pm0.02$\tabularnewline
$\chi_{\nu}^{2}\,(\mathrm{degrees\,of\,freedom})$ & $0.62\,(9455)$\tabularnewline
\hline 
\end{tabular}
\par\end{centering}

$^{a}$ Unabsorbed bolometric flux in the $0.01-100\,\mathrm{keV}$
band for \emph{Chandra} spectrum 17543
\end{table}
 The quality of the fit is similar to the previous fit (Table~\ref{tab:Fit-all-simpl}).
The largest difference is a lower $kT$: \texttt{simpl} rolls off
at the lower energies, and the blackbody moves to lower energies to
compensate. In turn, the flux is now underpredicted around $\sim3\,\mathrm{keV}$
(third panel of Figure~\ref{fig:The-persistent-spectrum}). If Comptonization
of the blackbody is the correct interpretation of the power law at
higher energies, the excess at lower energies must be produced by
another process. Photoionized reflection off the accretion disk could
produce this in combination with the Fe~K$\alpha$ line near $6.4\,\mathrm{keV}$
and the Compton hump.

\subsection{Relxill Reflection Model\label{sub:Relxill-Reflection-Model}}

We test the reflection interpretation using the \texttt{relxill} model
(Section~\ref{sub:Models-of-Photoionized}), with the complete \textsc{XSPEC}
model being \texttt{constant{*}TBabs(bbodyrad + cflux{*}relxill)}.
It is a complex model that, when fit to data of modest quality, presents
multiple degenerate solutions. For example, there are eight parameters
that shape the Fe~K$\alpha$ line. Moreover, several potentially
important effects are not taken into account, such as the dependence
on the density of the disk and the low-energy turn-off of the illuminating
power law \citep[fixed at $0.1$~keV;][]{Garcia2013}. In fitting this
model, one runs the risk of certain parameters taking on unphysical
values in order to compensate for these deficiencies. Indeed, when
left unconstrained, the fit prefers a maximally spinning neutron star
($a=1$) and an iron abundance of the disk of 10 times solar (the
maximal value provided by the model; see also \citealt{Degenaar2016}).
Therefore, we fix several parameters to reasonable values. We assume
a solar iron abundance. The disk emissivity is taken to decrease with
the third power of the radius, and the disk's outer radius is fixed
at a large value of $400\,R_{\mathrm{g}}$. As no eclipses or dipping
are apparent in the light curves, the disk's inclination is likely
less than $\sim60^{\circ}$: we choose a value of $30^{\circ}$. Similarly,
the fastest spinning neutron star known in an LMXB has spin $a\simeq0.3$
\citep[e.g.,][]{Degenaar2015}: we choose a value of $a=0.15$. During
the fits, the high energy cutoff pegs at the domain boundary of $1000\,\mathrm{keV}$.
Therefore, we fix the cutoff energy to this value. Furthermore, the
mentioned dependencies of the reflection spectrum on density and low-energy
turn-off are strongest at $E\lesssim3\,\mathrm{keV}$ (Section~\ref{sub:Models-of-Photoionized}).
We therefore limit the fit to the \emph{NuSTAR} spectra in their full
$3.0-50.0\,\mathrm{keV}$ band.

\begin{table}
\caption{\label{tab:Fit-nustar_refl}Fit of the \emph{NuSTAR} Persistent Spectra
with a Blackbody and a Reflected Power Law.}

\begin{centering}
\begin{tabular}{cc}
\hline 
\texttt{bbodyrad} & \tabularnewline
$kT\,(\mathrm{keV)}$ & $0.48\pm0.02$\tabularnewline
$K_{\mathrm{bb}}\,(\mathrm{km/10\,kpc})^{2}$ & $44\pm14$\tabularnewline
\hline 
\texttt{relxill} & \tabularnewline
$\Gamma$ & $2.060\pm0.013$\tabularnewline
$\log\xi$ & $3.28\pm0.08$\tabularnewline
$f_{\mathrm{refl}}$ & $0.31\pm0.04$\tabularnewline
$R_{\mathrm{in}}\,(10^{2}\,R_{\mathrm{g}})$ & $2.1_{-1.3}^{+1.9p}$\tabularnewline
$F_{3-50\,\mathrm{keV}}(\mathrm{10^{-10}erg\,s^{-1}\,cm^{-2}})$$^{a}$ & $0.716\pm0.004$\tabularnewline
\hline 
\texttt{constant} & scaling relative to FPMA\tabularnewline
\emph{NuSTAR} FPMB & $0.980\pm0.006$\tabularnewline
\hline 
$\chi_{\nu}^{2}\,(\mathrm{degrees\,of\,freedom})$ & $0.97\,(974)$\tabularnewline
\hline 
\end{tabular}
\par\end{centering}

$^{a}$ Unabsorbed bolometric flux in the $3.0-50\,\mathrm{keV}$
\emph{NuSTAR} band of the relxill component of the FPMA spectrum
\end{table}
We perform a fit with these constraints (Table~\ref{tab:Fit-nustar_refl}),
which is largely consistent with the results from \citet{Degenaar2016}.
The blackbody parameters are consistent within $1\,\sigma$ with the
fit of the blackbody + power law model (Table~\ref{tab:Fit-all-po}),
whereas the power law's $\Gamma$ is $5\%$ smaller. The inner disk
radius is large: $R_{\mathrm{in}}\simeq210\,R_{\mathrm{g}}$. The
width of the Fe~K$\alpha$ line is influenced by $R_{\mathrm{in}}$
as well as the neutron star spin and the disk's inclination. Repeating
the fit for spin $0<a<0.3$ and for inclination angles up to $60^{\circ}$
always leads to similarly large $R_{\mathrm{in}}\gtrsim10^{2}\,R_{\mathrm{g}}$.
A local minimum in $\chi^{2}$ is present at $\log\xi\simeq1.6$ and
a global minimum at $\log\xi=3.28\pm0.08$. The latter indicates that
the reflecting material is highly ionized. The reflection model provides
a good description of the Fe~K$\alpha$ line (Figure~\ref{fig:The-persistent-spectrum}
bottom), but a minor part of the Compton hump remains visible in the
residuals; its precise peak energy depends on the density of the reflector
\citep{Garcia2016}. Although we only fit to the \emph{NuSTAR} spectra,
we also show the residuals of a Chandra spectrum with respect to the
best-fitting model (Figure~\ref{fig:The-persistent-spectrum} bottom).
The features near $1\,\mathrm{keV}$ remain visible. This part of
the spectrum is most sensitive to density, or alternatively these
features could be produced by a local warm absorber. We refer to \citet{Degenaar2016}
for an in-depth discussion of the line features.

\section{Analysis of Burst Emission}

\label{sec:Analysis-of-Burst}

The X-ray count rates of the burst observations (Figure~\ref{fig:lcv}
top) exhibit a power law decline in the initial $1.6\times10^{5}\,\mathrm{s}$
since the \emph{MAXI} trigger (allowing for a small offset between
the PC and WT mode due to the accuracy of the pile-up correction and
differences in the grade selection). The burst from 2012 exhibited
strong variability from $6$ to $16$ minutes after the BAT trigger
\citep{Degenaar2013}. No such variability is visible in the 2015
burst, although this may have been missed due to the sparse sampling
in the first few hours. After this period ($2$ days since the trigger),
the count rate drops sharply. Over the subsequent $1.4$ weeks, the
count rate slowly increases to a value of $\sim1.2\,\mathrm{c\,s^{-1}}$.
The magnitudes of the UVOT detections exhibit similar behavior (Figure~\ref{fig:lcv}
bottom), suggesting that both parts of the spectrum are powered by
the same emission source, either directly or after reprocessing. The
burst in 2012 was similarly detected in the optical \citep{Ivarsen2012GCN,Meehan2012GCN}.
Considering thermal emission from a neutron star undergoing a thermonuclear
burst, a blackbody that peaks in the X-rays at $1\,\mathrm{keV}$
would produce a magnitude difference of $\Delta M\simeq1.2$ between
the UVW2 and U filters (not including possible differences in reddening
between bands). No such offset is apparent between the magnitudes
of different filters, which indicates that the spectrum in the UV
regime is relatively flat, possibly due to reprocessing of the burst
emission by the disk \citep[e.g.,][]{Ballantyne2004models,Hynes2006}.

Similar to the analysis of the persistent emission, we fit the burst
spectra both with a phenomenological model and a reflection model.

\subsection{Phenomenological Spectral Fits\label{sub:Phenomenological-Spectral-Fits}}

\begin{figure}
\includegraphics{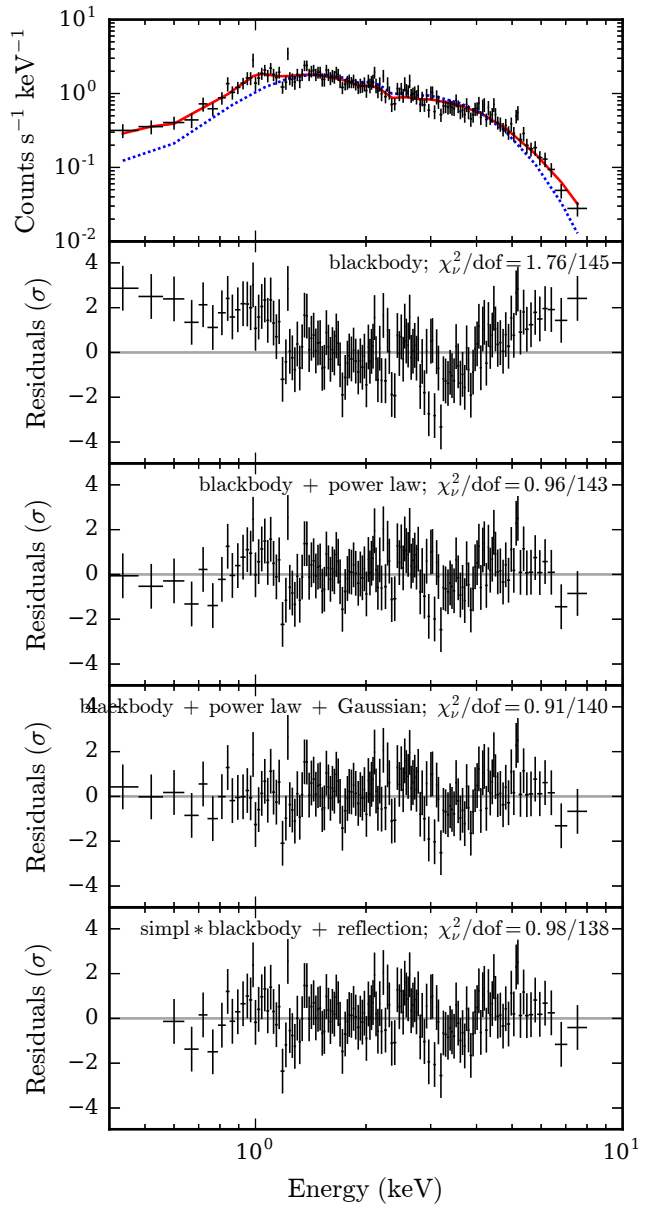}

\caption{\label{fig:s2_residuals}(Top) First XRT burst spectrum as a function
of energy. The dotted line is the best fit with an absorbed blackbody
model and the solid line results from a model that further includes
a power law and a Gaussian line (see also the second and fourth panels,
respectively). (Below) Residuals of spectral fits with indicated models,
goodness of fit, and degrees of freedom; in units of the $1\sigma$
uncertainty of the data points. A blackbody describes most of the
data, whereas a power law can fit the excesses at low and high energy.
The most prominent remaining feature in the residuals can be well
fit with a Gaussian emission line at $\sim1.0\,\mathrm{keV}$.}
\end{figure}
 
\begin{figure}
\includegraphics{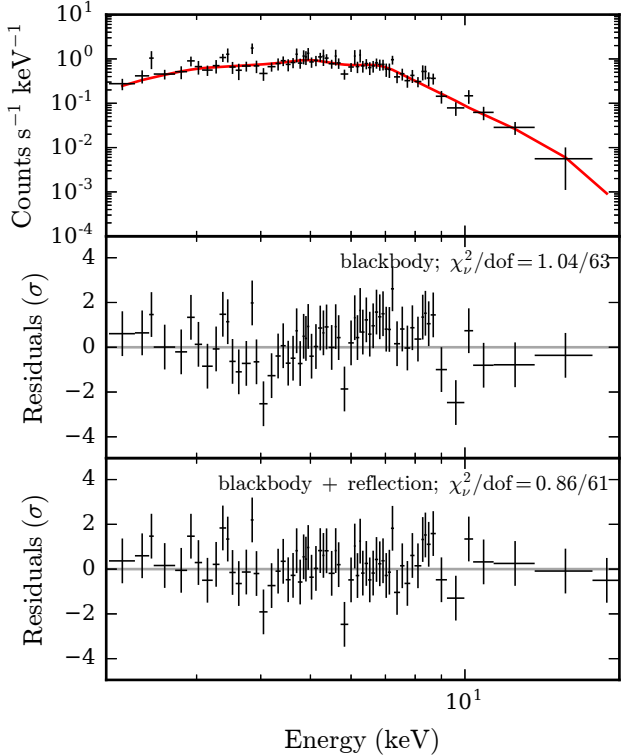}

\caption{\label{fig:s0_residuals}Similar to Figure~\ref{fig:s2_residuals}
for the first \emph{MAXI}/GSC burst spectrum. The solid line in the
top panel corresponds to the best fitting blackbody model that includes
a reflection component (see also the bottom panel). After a fit with
an absorbed blackbody, a broad excess around $\sim6.5\,\mathrm{keV}$
and a deficit above $\sim9\,\mathrm{keV}$ remain (middle panel).
These features are well described with a model of photoionized reflection
(bottom).}
\end{figure}
 We illustrate our choice of spectral model using the first XRT spectrum
(Figure~\ref{fig:s2_residuals}). The spectrum is dominated by a
blackbody \citep{Negoro2015,Iwakiri2015}, but excesses are visible
at both low and high energy (second panel of Figure~\ref{fig:s2_residuals}).
Adding a power law provides a reasonable description, although some
structure remains in the residuals (third panel of Figure~\ref{fig:s2_residuals}).
The strongest is an emission feature around $1\,\mathrm{keV}$ \citep[also present in the 2012 burst and the persistent spectrum;][]{Degenaar2013,Degenaar2016}
which can be fit with a Gaussian profile (fourth panel of Figure~\ref{fig:s2_residuals}).
The complete model is, therefore, similar to the persistent model
in Section~\ref{sub:Phenomenological-fit} with the addition of the
Gaussian: \texttt{TBabs(bbodyrad + cflux{*}powerlaw + gaussian)},
where interstellar absorption is again implemented with a fixed $N_{\mathrm{H}}$
(Section~\ref{sub:Interstellar-Absorption}).

This model is fit to all XRT spectra. The \emph{MAXI} spectra, however,
do not cover the $1\,\mathrm{keV}$ line and do not require the power
law. Those two spectra are fit with only an absorbed blackbody (Figure~\ref{fig:s0_residuals}).
\begin{figure}
\includegraphics{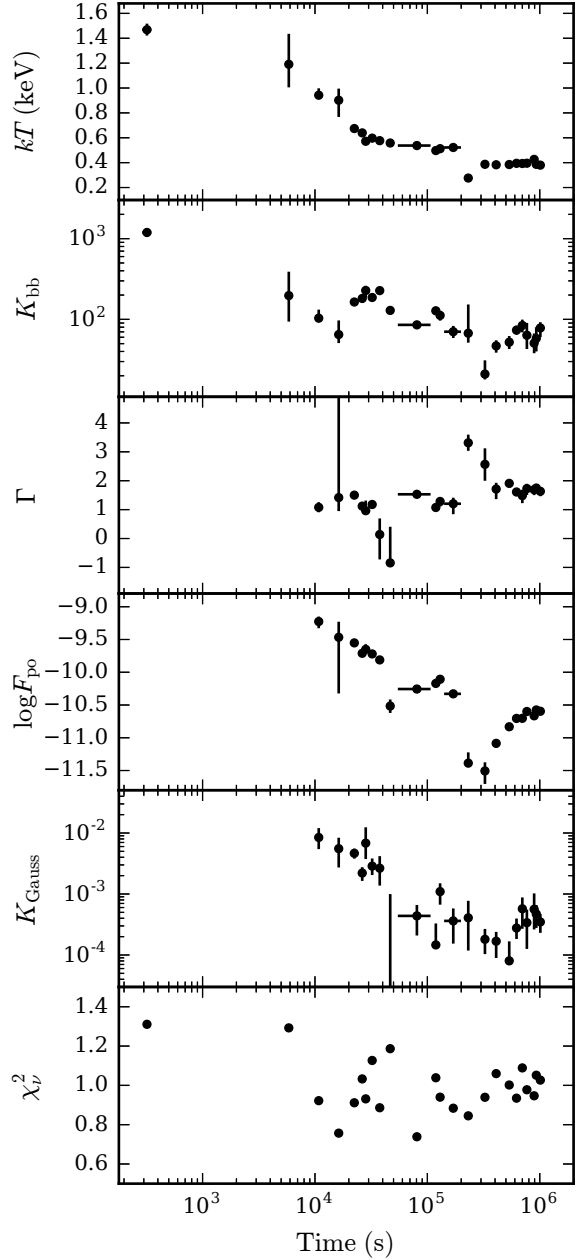}

\caption{\label{fig:po_fit}Best fit values and $1\sigma$ uncertainties as
a function of time from analysis of the \emph{MAXI} and XRT burst
spectra (using a time offset of $320\,\mathrm{s}$; see Section~\ref{sub:Burst-properties}).
The spectral components include a blackbody (temperature $kT$ and
normalization $K_{\mathrm{bb}}$), power law (photon index $\Gamma$
and $0.5-10\,\mathrm{keV}$ unabsorbed flux $F_{\mathrm{po}}$), and
a narrow Gaussian at $1\,\mathrm{keV}$ (normalization $K_{\mathrm{Gauss}}$).
The bottom panel shows the goodness of fit per degree of freedom,
$\chi_{\nu}^{2}$. Horizontal ``error bars'' indicate the width
of the time interval during which the spectrum of a data point was
observed. Normalizations are in units of $\mathrm{c\,s^{-1}\,cm^{-2}\,keV^{-1}}$.}
\end{figure}

\begin{figure}
\includegraphics{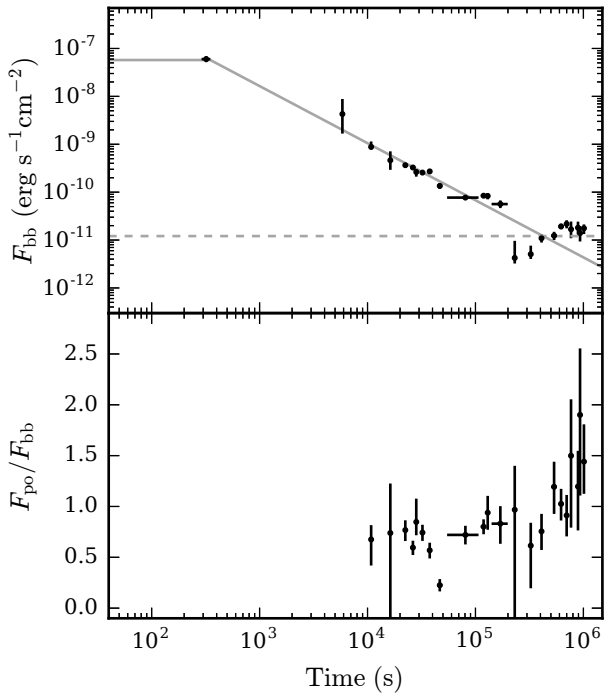}

\caption{\label{fig:flux_po}(Top) Unabsorbed bolometric flux from the blackbody
component, $F_{\mathrm{bb}}$, as a function of time since $320\,\mathrm{s}$
before the first data point. The solid line indicates the best-fit
model of the flux profile (Equation~\ref{eq:fit_function}) to the
first $4\times10^{4}\,\mathrm{s}$, and the dashed line indicates
the pre-burst blackbody flux (Table~\ref{tab:Fit-all-po}). (Bottom)
The ratio of the $0.5-10\,\mathrm{keV}$ power law flux, $F_{\mathrm{po}}$
(Figure~\ref{fig:po_fit}), and $F_{\mathrm{bb}}$. For the first
two data points $F_{\mathrm{po}}$ is not measured.}
\end{figure}
Both the power law and the blackbody are detected in all XRT spectra.
Even though the count rate is substantially lower at later times,
the spectra have similar statistics, as subsequent pointings were
combined (Table~\ref{tab:Observations}). The blackbody temperature
decreases from $kT=1.47\pm0.05\,\mathrm{keV}$ in the first (\emph{MAXI})
data point to a mean value of $kT=0.291\pm0.005\,\mathrm{keV}$ at
the end in a week of XRT pointings (Figure~\ref{fig:po_fit}). The
first data point indicates photospheric radius expansion (PRE), as
the blackbody normalization is an order of magnitude larger ($K_{\mathrm{bb}}=(1.2\pm0.2)\times10^{3}\,(\mathrm{km/10\,kpc})^{2}$)
than the mean value for the subsequent seven spectra ($K_{\mathrm{bb}}=(1.69\pm0.08)\times10^{2}\,(\mathrm{km/10\,kpc})^{2}$).
PRE is often observed around the peak of the brightest bursts, when
the flux reaches the Eddington limit \citep[e.g.,][]{Kuulkers2003}.
We further calculate the bolometric unabsorbed blackbody flux using
its proportionality to the the temperature, $F_{\mathrm{bb}}\propto\sigma T^{4}$,
and taking into account $K_{\mathrm{bb}}$ (Figure~\ref{fig:flux_po}
top). 

The power law index evolves over time: the first six XRT spectra yield
a weighted mean of $\Gamma=1.31\pm0.05$, and we find $\Gamma=1.72\pm0.06$
for the last eight spectra. In between, $\Gamma$ displays some variability.
We determine the unabsorbed in-band ($0.5-10\,\mathrm{keV}$) power
law flux, $F_{\mathrm{po}}$, and take the ratio to the blackbody
flux (on average $95\%$ of the bolometric blackbody flux falls in
the $0.5-10\,\mathrm{keV}$ band; Figure~\ref{fig:flux_po} bottom).
In the first six XRT spectra the weighted mean is $0.66\pm0.04$.
The ratio increases over time: for the last eight spectra it is $1.01\pm0.09$.
No power law component is detected in the first \emph{MAXI} scan,
but including a power law with the persistent value for $\Gamma$
(Table~\ref{tab:Fit-all-po}) one finds a $90\%$ upper limit of
$F_{\mathrm{po}}\lesssim1.0\times10^{-8}\,\mathrm{erg\,s^{-1}\,cm^{-2}}$,
which places an upper limit on $F_{\mathrm{po}}/F_{\mathrm{bb}}\lesssim0.17\pm0.02$. 

The Gaussian emission feature near $1\,\mathrm{keV}$ is outside of
the \emph{MAXI}/GSC band, but it is a significant component for the
first seven XRT spectra. The weighted mean of the centroid energy
is $1.035\pm0.009\,\mathrm{keV}$. The width of the Gaussian is small:
typically $\sim10^{-2}\,\mathrm{keV}$, which is consistent with being
unresolved. We investigate the significance of the line by comparing
the normalization, $K_{\mathrm{Gauss}}$, to its $1\sigma$ error:
for the first seven XRT spectra the mean value is $K_{\mathrm{Gauss}}=3.3\sigma$.
Furthermore, $K_{\mathrm{Gauss}}$ decreases at the same rate as the
blackbody flux.

\subsection{Reflection Fits\label{sub:Reflection-Fits}}

The phenomenological fits suggest that the burst flux, as represented
by the blackbody that dominates the spectrum, is reprocessed into
the power law and Gaussian components, whose fluxes follow the blackbody
flux. The flux fraction (Figure~\ref{fig:flux_po}) and the power
law's photon index (Figure~\ref{fig:po_fit}) are different in the
early and the later observations. This may indicate that two reprocessing
regions are active, and their relative contributions to the flux change
with time. Similar to our model of the persistent emission, we replace
the power law by a \texttt{simpl} component for Comptonization of
the blackbody emission (Section~\ref{sub:Simpl-Comptonization-Model})
and a reflection component. Because the burst emission is dominated
by the blackbody, we use a model of a reflected blackbody, which has
the same temperature as the blackbody component (\ref{sub:Models-of-Photoionized}).
The full model is: \texttt{TBabs(simpl{*}bbodyrad+cflux{*}relconv{*}reflection)}. 

After we replace the power law with these two components, competition
between them during the fit yields large uncertainties in the model
parameters. We therefore fit all burst spectra simultaneously and
constrain certain parameters to be the same for each spectrum, similar
to the procedure followed by, e.g., \citet{Keek2014sb2}. Specifically,
each spectrum has its own value of the blackbody parameters $kT$
and $K_{\mathrm{bb}}$ and the unabsorbed bolometric reflection flux
$F_{\mathrm{refl}}$ (determined in the $0.01-100\,\mathrm{keV}$
band), whereas the spectra share the ionization parameter $\log\xi$,
the inner radius of the reflection site $R_{\mathrm{in}}$, as well
as $\Gamma$ and $f_{\mathrm{sc}}$ of the \texttt{simpl} component.
Parameters of the \texttt{relconv} component (other than $R_{\mathrm{in}}$)
are the same as for the persistent emission (Section~\ref{sub:Relxill-Reflection-Model}).

\begin{table}
\caption{\label{tab:Fit-burst_refl}Fit of the Burst Spectra with a Comptonized
Blackbody and Reflection.$^{a}$}

\begin{centering}
\begin{tabular}{ccc}
\hline 
 & \emph{MAXI}/GSC & \emph{Swift}/XRT\tabularnewline
\hline 
\texttt{blackbody reflection} &  & \tabularnewline
$\log\xi$ & $2.7_{-0.2}^{+0.4}$ & $3.0_{-0.2}^{+0\mathrm{p}}$\tabularnewline
\hline 
\texttt{relconv} &  & \tabularnewline
$R_{\mathrm{in}}\,(R_{\mathrm{g}})$ & $(4.0_{-3.0}^{+0\mathrm{p}})\times10^{2}$ & $14_{-7}^{+25}$\tabularnewline
\hline 
\texttt{simpl} &  & \tabularnewline
$\Gamma$ & \textemdash{} & $2.3_{-0.3}^{+0.2}$\tabularnewline
$f_{\mathrm{sc}}$ & \textemdash{} & $0.56_{-0.10}^{+0.12}$\tabularnewline
\hline 
$\chi_{\nu}^{2}\,(\mathrm{degrees\,of\,freedom})$ & $0.86\,(61)$ & $1.04\,(1296)$\tabularnewline
\hline 
\end{tabular}
\par\end{centering}

$^{a}$ A subsection of the fit parameters is listed here for the
first \emph{MAXI} scan and the first six XRT spectra, where reflection
is detected. The XRT spectra are fit simultaneously, and the listed
values are shared between the spectra. The fit parameters that are
allowed to vary for each spectrum are shown in Figure~\ref{fig:refl_fit}.
`p' indicates where a fit parameter was pegged to the domain boundary.
\end{table}
 
\begin{figure}
\includegraphics{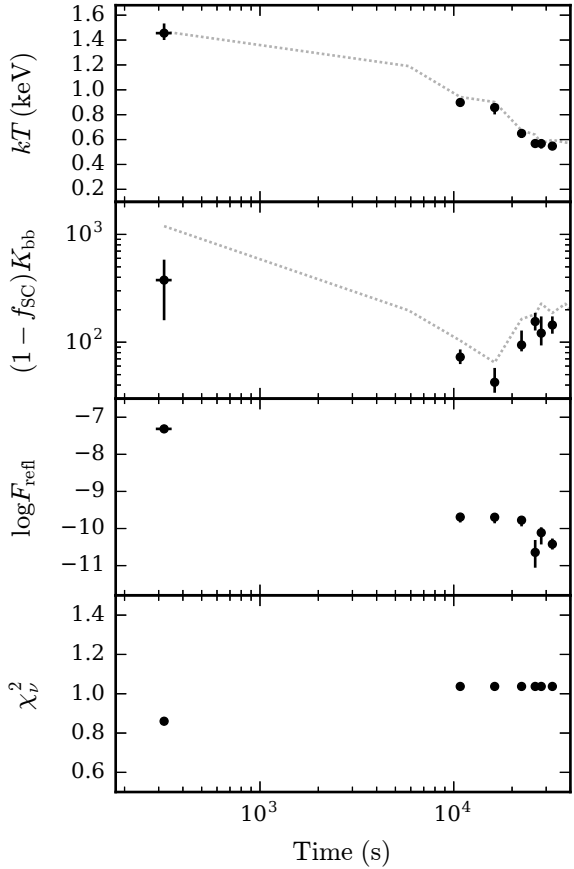}

\caption{\label{fig:refl_fit}Similar to Figure~\ref{fig:po_fit} for the
spectral fits with a reflection model. The spectral components include
a blackbody (temperature $kT$ and normalization $K_{\mathrm{bb}}$)
and a reflected blackbody (disk ionization parameter $\log\xi$ and
unabsorbed bolometric flux $F_{\mathrm{refl}}$). Part ($f_{\mathrm{sc}}$)
of the blackbody is Comptonized, such that the normalization of the
pure blackbody is $(1-f_{\mathrm{sc}})K_{\mathrm{bb}}$. Dotted lines
indicate values from the blackbody fit (Figure~\ref{fig:po_fit}).}
\end{figure}
 We fit the model to the XRT spectra, and find that after the first
six spectra, the reflection component can no longer be distinguished.
Therefore, we perform the simultaneous fit to the first six XRT spectra,
which cover the initial $9$ hours of the burst. We obtain a good
fit with $\chi_{\nu}^{2}=1.04$ (Table~\ref{tab:Fit-burst_refl}
and Figure~\ref{fig:refl_fit}). The fit residuals are similar to
those for the phenomenological model (Figure~\ref{fig:s0_residuals}
bottom).

Of the \emph{MAXI} spectra, only the first has sufficient counts to
distinguish deviations from a pure blackbody. No power law component
was found with the phenomenological spectral model (Figure~\ref{fig:s0_residuals}
second panel), and a fit with the reflection model finds a vanishingly
small $f_{\mathrm{sc}}$. Therefore, we fit the first \emph{MAXI}
spectrum without the \texttt{simpl} component (Table~\ref{tab:Fit-burst_refl}
and Figure~\ref{fig:refl_fit}). The modeled energy response of \emph{MAXI}/GSC
was tested in-orbit with observations of the Crab nebula \citep{Sugizaki2011maxiGSC}.
The Crab spectra indicate unmodeled features in the energy response,
but these deviations are smaller than the statistical error of our
spectra, and they are at different energies than the residuals of
a blackbody fit to the burst spectrum (Figure~\ref{fig:s0_residuals}
second panel). Those residuals are successfully described by a reflection
component (Figure~\ref{fig:s0_residuals} bottom).

\begin{figure}
\includegraphics{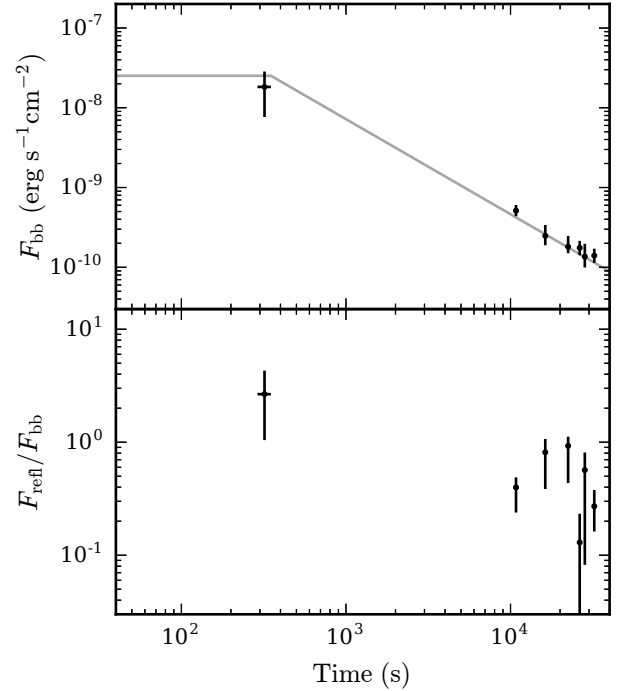}

\caption{\label{fig:flux_refl}Similar to Figure~\ref{fig:flux_po} for the
spectral fits with a reflection model (Figure~\ref{fig:refl_fit}).
The bottom panel shows the ratio of the unabsorbed bolometric fluxes
of the reflection ($F_{\mathrm{refl}}$) to the blackbody ($F_{\mathrm{bb}}$)
component. The solid line is the burst profile fit to the results
from the phenomenological fit multiplied by $(1-f_{\mathrm{sc}})$
(Figure~\ref{fig:flux_po}).}
\end{figure}
Compared to the phenomenological fits, $kT$ is on average $6\%$
smaller. $K_{\mathrm{bb}}$ is $55\%$ larger, but part of the blackbody
flux is Comptonized by \texttt{simpl}. The pure blackbody normalization
is $(1-f_{\mathrm{sc}})K_{\mathrm{bb}}$, which is $32\%$ smaller
than the phenomenological $K_{\mathrm{bb}}$. The changes in $kT$
and $K_{\mathrm{bb}}$ partially compensate each other, such that
the blackbody flux is lower by a factor $(1-f_{\mathrm{sc}})$ (Figure~\ref{fig:flux_refl}).
The reflection fraction, $F_{\mathrm{refl}}/F_{\mathrm{bb}}$, has
a weighted mean of $0.39\pm0.05$ for the XRT spectra. Interestingly,
the flux of the first \emph{MAXI} spectrum is also consistent with
being lower by a factor $(1-f_{\mathrm{sc}})$. Its reflection fraction
may, however, be substantially larger: $F_{\mathrm{refl}}/F_{\mathrm{bb}}=3\pm2$.

\subsection{Burst Properties}

\label{sub:Burst-properties}

To determine the properties of the burst, we consider the first $10.5\,\mathrm{hr}$,
because for later spectra the power law flux does not show smooth
behavior (Figure~\ref{fig:po_fit}), indicating that other effects
besides cooling start playing a role. This time interval includes
the two \emph{MAXI} spectra and the first seven XRT spectra. We use
the blackbody flux from the phenomenological fits to determine the
burst properties (see also discussion in Section~\ref{sub:Reprocessing-and-Anisotropic}).

The unabsorbed bolometric peak blackbody flux is $(6.0\pm0.9)\times10^{-8}\,\mathrm{erg\,cm^{-2}s^{-1}}$
in the first \emph{MAXI} scan, and the blackbody normalization is
substantially larger than in subsequent measurements. It is likely
an episode of PRE where the Eddington limit is reached. Because of
the PRE phase, we expect that the burst had a fast rise to the Eddington
limit \citep[$\lesssim 1\,\mathrm{s}$; e.g.,][]{Zand2014a}, and that
the flux stayed at this value until the end of PRE at time $t_{\mathrm{PRE}}$,
followed by a power law decay with index $\alpha$ produced by the
cooling neutron star envelope. This leads to the following simple
model of the burst flux as a function of time: 
\begin{equation}
\begin{array}{l}
F(t<t_{\mathrm{PRE}})=F_{\mathrm{Edd}}\\
F(t>t_{\mathrm{PRE}})=F_{\mathrm{Edd}}\left(\frac{t}{t_{\mathrm{PRE}}}\right)^{-\alpha}
\end{array}\label{eq:fit_function}
\end{equation}
 We fit this model to the burst flux. Because the first \emph{MAXI}
scan exhibits PRE and the second does not, we constrain $t_{\mathrm{PRE}}$
to lie between those two. The start time of the burst is unknown.
We therefore take a time offset, $t_{0}$, into account with respect
to the center of the first \emph{MAXI} scan. $t_{0}=30\,\mathrm{s}$
places the burst start at the beginning of the \emph{MAXI} scan, and
$t_{0}=92$~minutes places it at the start of the data gap preceding
the scan. We perform the fit for a linear grid of $1000$ values for
$t_{0}$ in this range. The effect of $t_{0}$ is largest for the
first data point: the fit is worse for increasing $t_{0}$, because
for large $t_{0}$ the first data point lies above the power law implied
by the other points. For each value of $t_{0}$ we determine the likelihood
of the fit by calculating the probability of obtaining the measured
$\chi^{2}$ or larger for $6$ degrees of freedom. Fits with $t_{0}>6.1\times10^{2}\,\mathrm{s}$
are disfavored at a $95\%$ confidence level: the burst start was
likely at most $10$ minutes before the first \emph{MAXI} scan.

For each fit we determine the optimal values of the parameters and
their $1\,\sigma$ positive and negative errors. We represent these
by distributions that peak at the optimal value and have Gaussian
``wings'' to lower and higher values as appropriate for the asymmetric
errors. The distributions are scaled by the likelihood of each fit,
and then summed for a given parameter over all fits. This produces
a distribution that includes the uncertainty in $t_{0}$ as well as
the likelihood of each value of $t_{0}$. It typically has a bell
shape with asymmetric wings. We report the value where the distribution
peaks as the optimal value, and integrate the wings outwards to locate
the $68\%$ confidence regions. We find $\alpha=1.15{}_{-0.12}^{+0.14}$,
$F_{\mathrm{Edd}}=(5.7_{-0.7}^{+1.0})\times10^{-8}\,\mathrm{erg\,s^{-1}\,cm^{-2}}$,
and $t_{\mathrm{PRE}}=(1.7{}_{-1.1\mathrm{p}}^{+1.7})\times10^{2}\,\mathrm{s}$,
where `p' indicates for $t_{\mathrm{PRE}}$ that the lower error is
pegged to the end of the first \emph{MAXI} scan. $F_{\mathrm{Edd}}$
is slightly lower than the flux value of the first data point, because
a lower $F_{\mathrm{Edd}}$ is preferred for larger values of $t_{0}$.
The difference is, however, only $0.3\,\sigma$. For each fit we calculate
the burst fluence, $\mathcal{F}_{\mathrm{burst}}$, and the ratio
of the fluence and peak flux, $\tau$, and combine the results in
the same way as the fit parameters: $\mathrm{\mathcal{F}_{\mathrm{burst}}=(7.2_{-1.3}^{+2.2})\times10^{-5}\,}\mathrm{erg\,cm^{-2}}$
and $\tau=(1.3_{-0.3}^{+0.4})\times10^{3}\,\mathrm{s}$, where the
fluence is calculated from $t=0$ to the end of the ninth data point
where we ended the fit.

Now that we have a description of the flux decay, we check whether
the time resolution that we employ is sufficient. During the exposure
interval of each spectrum, we compare the expected change in the blackbody
temperature from the central value (half of the total change), $\Delta kT$,
to the uncertainty, $\sigma$, in the measured $kT$ for the respective
spectrum. For the majority of spectra, $\Delta kT\lesssim1\,\sigma$,
and for one spectrum $\Delta kT=3.7\,\sigma$. Therefore, for all
but the latter case the change in $kT$ is small enough to allow for
fitting with a single temperature blackbody. The worst case is the
spectrum just before $t=10^{5}\,\mathrm{s}$, which spans a substantially
longer time interval than the other spectra. Because this spectrum
was taken after the first $10.5\,\mathrm{hr}$ where we determine
the burst properties, and because the spectral parameters do not exhibit
especially deviant behavior, we forego splitting the spectrum in shorter
time intervals.

\section{Discussion}

\label{sec:Discussion}

We first investigate the properties of the burst and persistent emission
using the phenomenological fits. Next, we use the results of the spectral
analysis to form a picture of the accretion environment in \source{},
and discuss the implications of reflection and anisotropic emission
on the burst properties.

\subsection{Comparison to Other Long X-Ray Bursts}

\label{sub:Comparison-to-other}

A comparison of the 2012 and 2015 bursts from \source{} is challenging
because of the different coverage and data quality of the two bursts.
The 2012 burst was first detected by \emph{Swift}/BAT, and the BAT
spectrum yields a blackbody flux of $4.8_{-1.6}^{+2.8}\times10^{-8}\,\mathrm{erg\,s^{-1}cm^{-2}}$
\citep{Degenaar2013}, which is $0.4\sigma$ smaller than the peak
flux that we derive for the 2015 burst (Section~\ref{sub:Burst-properties}).
For the 2015 burst we find evidence of PRE, and one expects the same
for the 2012 burst, because the flux variations in its tail are thought
to be associated with strong expansion at the onset \citep{Degenaar2013}.
Nevertheless, PRE was not detected for the 2012 burst, likely because
of the limited data quality. The BAT count rate peaked halfway through
the exposure, and \citet{Degenaar2013} interpreted this as the true
maximum in the burst flux, arguing that the peak flux was larger than
the time-averaged BAT flux. However, because of the BAT's energy band
($>15\,\mathrm{keV}$), it is less sensitive to the PRE phase, where
the blackbody temperature is reduced. The BAT spectrum was, therefore,
dominated by the post-PRE phase, and the peak in the BAT count rate
may coincide with the end of the PRE phase (``touchdown''; e.g.,
compare the count rate in different bands for a burst from SLX~1735-269
in \citealt{Molkov2005}). We suggest that the flux measured from
the BAT spectrum was close to the Eddington limit, and both the 2012
and 2015 bursts reached this limit.

If the peak in the BAT count rate coincides with the end of the PRE
phase, the 2012 burst had a PRE duration of at least $t_{\mathrm{PRE}}\gtrsim80\,\mathrm{s}$
(the time from the burst start up to the BAT trigger is not included),
which is compatible with the $t_{\mathrm{PRE}}=(1.7{}_{-1.1\mathrm{p}}^{+1.7})\times10^{2}\,\mathrm{s}$
that we infer for the 2015 burst (Section~\ref{sub:Burst-properties}).
$t_{\mathrm{PRE}}$ is typically proportional to the burst duration
\citep[e.g.,][]{Zand2010}, and based on $t_{\mathrm{PRE}}$ the 2015
burst is of similar or substantially longer duration than the 2012
burst. Only the early part of the 2012 burst was observed, and the
presence of strong flux variability makes it challenging to constrain
the cooling trend. \citet{Degenaar2013} suggest a linear trend, which
implies a much faster flux decay than the expected power law \citep[e.g.,][]{Zand2014}.
Interestingly, robotic optical telescopes detected a declining R-band
flux in the $2$ hours following the \emph{Swift}/BAT trigger \citep{Ivarsen2012GCN,Meehan2012GCN},
and from $1.1\times10^{3}\,\mathrm{s}$ to the end of the observations
at $7.0\times10^{3}\,\mathrm{s}$ the flux decay follows a power-law
trend. Perhaps the burst tail continued longer than the previously
inferred duration of $1.1\times10^{3}\,\mathrm{s}$ \citep{Degenaar2013}.
Furthermore, the reported burst fluence is $\sim1/4$ of the 2015
burst \citep{Degenaar2013}, but it may have been underestimated as
well, if the unobserved tail continued for longer than previously
assumed.

Comparing the 2015 burst to bursts from other sources, it is observable
for hours, similar to superbursts \citep[e.g.,][]{Keek2008int..work}.
Therefore, the burst was initially tentatively classified as a superburst
\citep{Negoro2015,Iwakiri2015}. However, the day-long duration is
in part due to the low persistent flux, which allows the burst to
be detectable for many hours \citep[see][for a similar event lasting 3 hours]{Degenaar2011}.
A better quantity for comparison is the decay time scale $\tau$,
which is derived from the burst fluence (Section~\ref{sub:Burst-properties}).
We find $\tau=21\pm7$~minutes for the 2015 burst, whereas superbursts
typically have $\tau\gtrsim69$~minutes \citep{Strohmayer2002,Keek2008int..work}.
It is, therefore, more likely to be a long helium burst, i.e. an intermediate
duration burst. It is among the most energetic helium flashes observed
thus far \citep[e.g.,][]{Linares2012}. Some of the longer specimens
from other sources have $\tau\simeq4$~minutes \citep[XMMU J174716.1-281048;][]{Degenaar2011},
$\tau\simeq7$~minutes \citep[SLX 1735-269;][]{Molkov2005}, $\tau\simeq8$~minutes
\citep[SLX 1737-282;][]{2002JZ}, and $\tau\simeq16$~minutes and
$\tau\simeq43$~minutes \citep[4U 1850-086;][based on the reported exponential decay timescales]{Zand2014ATel,Serino2016}.
A common issue for these bursts is again sparse coverage and modest
data quality, which makes the precise measurement of the burst properties
challenging. Nevertheless, one long burst from 4U~1850-086 exhibited
$t_{\mathrm{PRE}}=(4.8\pm0.5)\times10^{2}\,\mathrm{s}$ \citep{Zand2014ATel},
suggesting that it was a more energetic event than the 2015 burst
from \source{}. Furthermore, for several superbursts from other sources
with low mass accretion rates, there remains ambiguity as to whether
the fuel consists of carbon or helium, although their interpretation
as carbon flashes remains preferred \citep[e.g.,][]{Kuulkers2010,Altamirano2012,Serino2016}.

\subsection{Burst Ignition\label{sub:Burst}}

We regard the flux of the first \emph{MAXI} scan as the Eddington
limited flux. Equating the luminosity to the empirical Eddington luminosity
of $L_{\mathrm{Edd}}=3.8\times10^{38}\,\mathrm{erg\,s^{-1}}$ \citep{Kuulkers2003},
we find a distance of $d=7.3\pm0.5\,\mathrm{kpc}$. \citet{Degenaar2013}
derived a distance of $5\,\mathrm{kpc}$ by extrapolating the peak
flux of the 2012 burst, but we suggested that this may not be the
correct measure of the Eddington flux (Section~\ref{sub:Comparison-to-other}).

Using the above distance, we find the burst energy of $E_{\mathrm{b}}=(4.5_{-1.0}^{+1.6})\times10^{41}\,\mathrm{erg}$
in the observer frame. Part of the burst energy may escape the neutron
star as a neutrino flux. We neglect it here, as it is expected to
be minor for columns of $y\lesssim10^{11}\,\mathrm{g\,cm^{-2}}$ \citep{Keek2011}.
During PRE, a substantial part of the burst energy may power a radiative
wind \citep[e.g.,][]{Weinberg2006}. The fraction of the fluence during
the PRE phase is $t_{\mathrm{PRE}}F_{\mathrm{Edd}}/\mathcal{F_{\mathrm{burst}}}=0.13_{-0.10}^{+0.14}$.
The total burst energy could, therefore, be larger by $\sim10\%$.
To calculate the ignition column depth, we take an energy release
of burning He to Ni of $E_{\mathrm{nuc}}\simeq10^{18}\,\mathrm{erg\,g^{-1}}$,
and assume a neutron star mass of $1.4\,M_{\odot}$ and radius of
$R=10\,\mathrm{km}$, which give a gravitational redshift of $1+z=1.31$.
We find 
\[
y_{\mathrm{ign}}=\frac{(1+z)E_{\mathrm{b}}}{4\pi R^{2}E\mathrm{_{nuc}}}=(4.7_{-1.0}^{+1.6})\times10^{10}\,\mathrm{g\,cm^{-2}}.
\]

\subsection{Persistent Accretion\label{sub:Persistent}}

We have fit several models to the persistent flux (Section~\ref{sec:pers_spectra}).
When extrapolating the model to lower energies, a power law gives
an unphysically large bolometric flux. The reflection model has a
low-energy turn-over of the power law at $0.1\,\mathrm{keV}$, but
the turn-over is likely near the peak of the blackbody at an order
of magnitude higher energy. Therefore, we prefer the bolometric flux
obtained with the \texttt{simpl} model (Table~\ref{tab:Fit-all-simpl}).
Multiplying the flux measured for \emph{Chandra} spectrum 17543 by
the \emph{MAXI} scaling factor, we find a bolometric unabsorbed flux
of $F_{\mathrm{pers}}=(0.92\pm0.07)\times10^{-10}\mathrm{erg\,s^{-1}\,cm^{-2}}$.
This is the time-averaged persistent flux between the bursts in 2012
and 2015. Using the flux of the first \emph{MAXI} burst observation
as a measure of the Eddington limit, we find $F_{\mathrm{pers}}/F_{\mathrm{Edd}}=(1.5\pm0.3)\times10^{-3}$. 

Only a few LMXBs have shown bursts at such low persistent flux \citep[e.g.,][]{Kaptein2000,Degenaar2010,Chenevez2012ATel},
and \source{} presents a rare instance of recurring bursts at low
mass accretion rate \citep[1RXS J171824.2-402934 is another example;][]{Zand2009J1718}.
The time between the triggers of the bursts in 2012 and 2015 is $3.3552\,\mathrm{yr}$,
and the ratio of the persistent and burst fluences is $\alpha=\mathcal{F}_{\mathrm{pers}}/\mathcal{F}_{\mathrm{burst}}=(1.4\pm0.4)\times10^{2}$.
Assuming $100\%$ efficiency of converting gravitational energy to
X-rays, a column of $y_{\mathrm{acc}}=(2.7\pm0.4)\times10^{10}\,\mathrm{g\,cm^{-2}}\left(\frac{1.4\,M_{\odot}}{M}\right)\left(\frac{10\,\mathrm{km}}{R}\right)$
was accreted. This includes the uncertainty in the distance. When
comparing $y_{\mathrm{acc}}$ to $y_{\mathrm{ign}}$, we can avoid
this uncertainty by simply taking
\[
\frac{y_{\mathrm{acc}}}{y_{\mathrm{ign}}}=\frac{\mathcal{F}_{\mathrm{pers}}}{\mathcal{F}_{\mathrm{burst}}}\frac{E_{\mathrm{nuc}}}{GM/R}=0.73_{-0.14}^{+0.23}\left(\frac{1.4\,M_{\odot}}{M}\right)\left(\frac{R}{10\,\mathrm{km}}\right),
\]
 which means that the two columns are consistent within $1.2\,\sigma$.
This confirms that the X-ray luminosity is indeed a good measure of
the mass accretion rate in this system. Unfortunately, there were
many instances when the source was not observed by any X-ray instrument
for several hours, and we cannot verify that indeed no other burst
occurred between 2012 and 2015. \emph{MAXI} covered the source with
at least one scan per 3 hours for $63\%$ of the time, and similarly
gaps are present in the coverage by \emph{Swift}/BAT.

$y_{\mathrm{ign}}$ is substantially smaller than typical for superbursts
\citep{Cumming2006}, especially at low mass accretion rates \citep[e.g.,][]{Kuulkers2010,Altamirano2012},
where carbon ignition is expected to occur much deeper \citep{Cumming2006,Keek2011}.
This comfirms our classification of the event as due to deep helium
ignition (Section~\ref{sub:Comparison-to-other}). Models of pure
helium accretors, however, predict $y_{\mathrm{ign}}$ to be much
larger at the observed mass accretion rate \citep{Cumming2006}. These
models consider heating of the neutron star envelope by nuclear processes
in the crust and neutrino cooling in the core. The maximum heating
that this provides is insufficient to explain the observed $y_{\mathrm{ign}}$,
and additional ``shallow'' heating would be required to ignite helium
at the observed depth \citep{Brown2009,Deibel2015,Deibel2016}. This
burst from \source{} could indicate that shallow heating is active
even at very low mass accretion rates. Alternatively, the accretion
composition could include some hydrogen, the burning of which could
heat the envelope and lower $y_{\mathrm{ign}}$ to the observed value.
At low mass accretion rates, hydrogen may burn at a small depth, producing
a deep layer of helium \citep[e.g.,][]{Fujimoto1981,Peng2007}, and
intermediate duration bursts have been observed from such sources
\citep[e.g.,][]{Falanga2009}. Optical spectroscopic observations
of the companion star could determine the presence of hydrogen.

\subsection{Reflection Signals}

\label{sub:Reflection-Signals}

The 2012 burst spectrum exhibited a strong emission line near $1\,\mathrm{keV}$,
which was suggested to be a fluorescent Fe~L line originating at
a distance of $\sim10^{2}\,R_{\mathrm{g}}$ from the neutron star
\citep{Degenaar2013}. We find an emission feature with similar energy
and width in the XRT spectra of the 2015 burst. The normalization
of the line decreases at the same rate as the blackbody flux, and
the line during the 2012 burst roughly follows the same trend. This
is a strong indication that the line is produced by reprocessed burst
emission. High resolution \emph{Chandra} spectra of the persistent
emission also exhibit emission features near $1\,\mathrm{keV}$. Furthermore,
the \emph{NuSTAR} persistent spectra clearly exhibit a broadened Fe~K$\alpha$
line and a Compton hump: the tell-tale signs of photoionized reflection
\citep{Degenaar2016}. Therefore, we argue that the $1\,\mathrm{keV}$
feature is similarly produced by reflection, and we have assumed reflection
to be dominated by a single reflection component at each moment in
time. During the bursts, the feature is accompanied by a soft excess,
which may be the Bremsstrahlung continuum from reflection \citep[e.g.,][]{Ballantyne2004models}.
Reflection models can describe the soft excess and part of the $1\,\mathrm{keV}$
line, but residuals remain around the line both for the 2015 burst
(Figure~\ref{fig:s2_residuals} bottom) and the persistent spectra
(Figure~\ref{fig:The-persistent-spectrum} bottom; see also \citealt{Degenaar2016}).
The strength of the lines in the reflection spectrum is highly dependent
on the composition \citep[e.g.,][]{Madej2014} and the density of
the reflecting material \citep{Ballantyne2004models}. Current reflection
models are limited in this respect, which may be the cause of the
imperfect description of the observed feature.

The reflection models suggest that the disk is highly ionized ($\log\xi\simeq3$)
both during the burst and the persistent spectrum, even though the
illuminating flux changes by orders of magnitude. During two superbursts,
$\log\xi$ was observed to decrease over time \citep{Ballantyne2004,Keek2014sb2},
but for several other sources the persistent $\log\xi$ is similarly
large \citep[e.g.,][]{Cackett2010,Degenaar2015,Iaria2016}. As $\xi$
depends on both flux and density (Section~\ref{sub:Models-of-Photoionized}),
one explanation could be an increase in the density of the reflection
site during the burst, for example if the burst strips the low-density
upper layer from the disk. Low density material could also be present
near the inner edge of a disk that is truncated by the neutron star's
magnetic field \citep{Ballantyne2012models}. The persistent flux
could reflect off this material, whereas the burst flux may reflect
off the inner disk. Another possibility is that the inner disk is
overionized, such that it does not produce emission lines. Instead
of the inner radius of the disk, $R_{\mathrm{in}}$ would correspond
to a location further out in the disk, where $\xi$ is sufficiently
reduced to produce the reflection features. For stronger illumination,
$R_{\mathrm{in}}$ would then increase, whereas we see it decrease
during the burst. Furthermore, overionization has not been inferred
for other LMXBs where the inner disk is similarly highly ionized \citep[e.g.,][]{Cackett2010}. 

Care must be taken, however, not to overinterpret the present results,
because the uncertainties in $R_{\mathrm{in}}$ are large, at least
in part due to the limitations of the reflection models (Section~\ref{sub:Models-of-Photoionized}).
Moreover, the fit parameters may be subject to small systematic changes
if we had employed reflection models for a hydrogen-rich accretion
disk, instead of a helium-rich disk (Section~\ref{sub:Models-of-Photoionized}).
More importantly, the soft reflection spectrum ($\lesssim2\,\mathrm{keV})$
depends strongly on the density of the reflector (Figure~\ref{fig:reflection_models}),
and fits with models assuming, e.g., a lower density would likely
yield a larger reflection fraction.

\subsection{Burst Impact on Accretion Environment}

\label{sub:Burst-Impact-on}

Both the persistent and burst spectra are well-described by a blackbody
and a power law. The blackbody likely originates at the neutron star
surface, which is heated by persistent accretion and by thermonuclear
burning during the burst. The blackbody temperature demonstrates cooling
in the burst tail, which is often cited as a defining characteristic
of a Type I X-ray burst of thermonuclear origin. Over the course of
a day, the blackbody flux decreases with time as $F_{\mathrm{bb}}\propto t^{-1.15\pm0.14}$.
The power is just $1.1\,\sigma$ smaller than the expected $\simeq1.3$
for bursts that ignite at large column depths where cooling is dominated
by ions \citep{Zand2014}.

During the burst and the subsequent $11$ days, the power law flux
traces the changes in the blackbody flux. The power law is likely
produced by reprocessing of the blackbody emission. We speculate that
this is Comptonization in an optically thin plasma near the neutron
star. We refer to it as a ``corona,'' but it may be a boundary layer
or accretion flow between a truncated disk and the neutron star. As
no high-energy cut-off is detected in the \emph{NuSTAR} band, the
electrons in the plasma must be hot: $kT_{\mathrm{e}}\gtrsim10^{2}\,\mathrm{keV}$.
During the PRE phase at the start of the burst, the \emph{MAXI} spectrum
shows no sign of the power law, suggesting disruption of the corona
by the burst, reducing its scattering fraction by at least a factor
$4$ (Section~\ref{sub:Phenomenological-Spectral-Fits}). In the
burst tail the corona may have reformed, since the power law returns
with a similar $\Gamma$ and a slightly lower scattering fraction
than in the persistent spectrum.

Fits with reflection models find an inner disk radius of $R_{\mathrm{in}}\sim10^{2}\,R_{\mathrm{g}}$
for the persistent emission, which suggests that the disk is truncated
far from the neutron star \citep{Degenaar2016}, whereas in the burst
tail the disk may have moved inwards to $R_{\mathrm{in}}\sim10^{1}\,R_{\mathrm{g}}$.
Possibly, in the tail Poynting-Robertson drag brings the inner disk
closer to the neutron star \citep{Walker1992,Ballantyne2005,Worpel2013,Worpel2015}.
At the burst start a large $R_{\mathrm{in}}$ is favored, and the
effect of Poynting-Robertson drag may have been temporarily kept at
bay by radiation pressure or an outflowing wind. However, the quality
of the burst spectra is modest, and the uncertainty in $R_{\mathrm{in}}$
is large. 

Another quantity to consider is the reflection fraction, which at
all times is substantial. For the persistent emission, a large $R_{\mathrm{in}}$
and reflection fraction may be consistent if the disk is illuminated
by a corona that extends at least to $R_{\mathrm{in}}$, such that
a substantial part of the power law flux is intercepted by the disk.
In the burst, the blackbody dominates the reflection signal instead
of the power law. In the tail, the measured reflection fraction meets
the expectations for a flat disk that extends to the neutron star
\citep{lapidus85mnras,fujimoto88apj,He2016}, and $R_{\mathrm{in}}$
at this time is indeed consistent with no or a small gap between star
and disk. At the burst start, however, the reflection fraction is
substantially larger than unity (with a large error). This may require
a steeply inclined reflection surface close to the star \citep{He2016},
but the measured $R_{\mathrm{in}}$ is large. However, the large uncertainty
of $R_{\mathrm{in}}$ allows for a value of similar size as in the
tail. Therefore, we favor the scenario where the disk is truncated
at $R_{\mathrm{in}}\sim10^{2}\,R_{\mathrm{g}}$ during persistent
accretion, whereas the impact of the burst brings the inner disk close
to the neutron star, both at the burst start and in the tail.

During the burst, the behavior of the source is dominated by the cooling
of the blackbody, until the total bolometric flux is reduced to the
pre-burst persistent level. Over the course of a week, the spectrum
returns to being dominated by the power law. At the end of the XRT
observations, both $\Gamma$ and the scattering fraction are lower
than the persistent values. Perhaps the corona has not fully recovered
from the burst, which may take longer than for other bursting LMXBs,
because of the low mass accretion rate. The influence of a burst on
the hard tail has previously been inferred for a superburst \citep{Keek2014sb1}
and by stacking short bursts \citep{Maccarone2003,Chen2013,Ji2014,Kajava2016}.
Interestingly, at $t\simeq2\times10^{5}\,\mathrm{s}$ the spectrum
suddenly changes, and it takes several days to return to its previous
behavior. The blackbody, which we attribute to the cooling neutron
star, is strongly suppressed at this time, and the overall source
flux is reduced. The flux reduction could be the result of an inner
disk depleted by Poynting-Robertson drag. The restoration of the inner
disk and the accretion flow might briefly block our view of the neutron
star.

The limitations posed by the modest quality of the burst spectra and
by the reflection models make it challenging to put firm constraints
on the interaction of the burst and the accretion environment. It
is however, clear that the burst has a strong influence on both the
``corona'' and the disk \citep[for further discussion, see][]{Ballantyne2005}.

\subsection{Reprocessing and Anisotropic Emission\label{sub:Reprocessing-and-Anisotropic}}

We have derived the burst properties under the assumption of isotropic
emission, which is not the case in LMXBs \citep{lapidus85mnras,fujimoto88apj,He2016}.
Now that we have a picture of the different emission and reprocessing
sites in \source{}, we investigate how this affects the observed
burst flux. Anisotropy factors for burst emission have only been calculated
for disks that extend to the neutron star surface. The burst tail
may be the closest to this situation. The measured reflection fraction
of $0.39\pm0.05$ corresponds to a geometrically thin disk with an
inclination angle of $39^{\circ}\pm7^{\circ}$ \citep{He2016}, which
is consistent with the absence of eclipses and dips, and is $1.3\,\sigma$
larger than the $30^{\circ}$ that we assumed in Section~\ref{sub:Relxill-Reflection-Model}.
It implies an anisotropy factor of $\xi_{\mathrm{d}}^{-1}=0.92\pm0.03$
for the direct blackbody emission \citep{He2016}, and the intrinsic
direct neutron star flux is larger by $9\%$ than observed. For truncated
disks, blocking of the line of sight by the disk is smaller, and $\xi_{\mathrm{d}}^{-1}$
is even closer to unity. 

We find that a scattering fraction $f_{\mathrm{sc}}$ of the blackbody
emission is reprocessed by a Comptonizing corona. If the corona has
a spherical geometry that envelops the neutron star, its anisotropy
is similarly small as for the blackbody itself. Part of the blackbody
is scattered from the line of sight, and the intrinsic neutron star
emission is the sum of the observed blackbody and the Comptonized
part. Indeed, the burst fit that includes the \texttt{simpl} model
returns a blackbody flux that is lower by a factor $(1-f_{\mathrm{sc}})$
than the blackbody flux from the phenomenological fit (Section~\ref{sub:Reflection-Fits}).\footnote{The power law in Section~\ref{sub:Phenomenological-Spectral-Fits}
fits mostly to the soft excess, which is described by reflection in
Section~\ref{sub:Reflection-Fits}, whereas the \texttt{simpl} component
fits to the hard tail.} The latter is, therefore, a good measure of the intrinsic isotropic
burst emission. The anisotropy correction for the persistent flux
may be similarly small.

At the start of the burst, we suspect that the disk may not be flat.
The anisotropy is strongly dependent on the disk geometry \citep{He2016},
and in principle a large correction could be needed: $\xi_{\mathrm{d}}^{-1}\sim0.1$.
This would have important consequences for the derived quantities
such as the distance and $y_{\mathrm{ign}}$. Fortunately, the Comptonizing
corona is absent at the burst start, and we can regard all flux to
be from the blackbody and its reflection, the sum of which may underestimate
the intrinsic burst emission by at most tens of percent \citep{He2016}.

\section{Conclusions and Outlook}

\label{sec:Conclusions-and-Outlook}

We performed a time-resolved spectral analysis of the 2015 Type I
X-ray burst from \source{} observed with \emph{MAXI} and \emph{Swift}
\citep{Negoro2015,Iwakiri2015}. The burst is of exceptional duration,
both because its deep ignition providing a long decay timescale of
$\tau\simeq21$ minutes, and because the low persistent flux of $(1.5\pm0.3)\times10^{-3}\,F_{\mathrm{Edd}}$
allows the burst cooling to be detectable for close to a day. Analysis
of the persistent flux observations with \emph{MAXI}, \emph{Chandra},
and \emph{NuSTAR} shows that the column accreted since the previous
burst in 2012 is consistent with the ignition column of $y_{\mathrm{ign}}=(4.7_{-1.0}^{+1.6})\times10^{10}\,\mathrm{g\,cm^{-2}}$
derived for the 2015 burst. The burst onset exhibits radius expansion,
and the tail describes a straight power law: $F\propto t^{-1.15\pm0.14}$.
We categorize this event as an intermediate duration burst from deep
helium burning.

Both the burst and persistent spectra are well described by a blackbody
and a power law \citep[see also][]{Degenaar2016}. The persistent
\emph{NuSTAR} spectra exhibit clear evidence of photoionized reflection
of the power law \citep{Degenaar2016}. The \emph{Swift}/XRT burst
spectra exhibit a soft excess and an emission line at $1\,\mathrm{keV}$
that also suggest reflection of the burst off the disk. We investigate
a description of the spectra that includes burst reflection and a
Comptonized component (instead of the power law). The Comptonized
part is missing at the peak of the burst, and has returned in the
tail: possibly a Comptonizing corona is temporarily disrupted in the
brightest phase of the burst. The reflection models find the disk
to be highly ionized at all times, but in our interpretation the reflection
location moves substantially closer to the neutron star during the
burst from $R_{\mathrm{in}}\simeq2\times10^{2}\,R_{\mathrm{g}}$ to
$\simeq14\,R_{\mathrm{g}}$. Poynting-Robertson drag exerted by the
burst could increase the inflow of matter \citep{Walker1992,Worpel2013,Worpel2015}.

The \emph{Neutron Star Interior Composition Explorer} \citep[NICER;][]{Gendreau2012NICER}
to be launched in 2017 will host a $17$ times larger effective area
at $1\,\mathrm{keV}$ than \emph{Swift}/XRT, whereas \emph{ATHENA}'s
\citep{BarconsATHENA} Wide Field Imager \citep[WFI;][]{Meidinger2014AthenaWFI}
promises a $131$ times larger effective area (launch in 2028). Future
observations of intermediate duration bursts with these instruments
will, therefore, provide a detailed view of the interesting processes
that we glimpsed with \emph{MAXI} and \emph{Swift}. Moreover, \emph{NICER},
\emph{ATHENA}, and a mission like the \emph{Large Observatory for
X-Ray Timing} \citep[LOFT;][]{Feroci2014LOFT,Zand2015LOFT} with a
collecting area of $\sim8\,\mathrm{m^{2}}$ will also be able to detect
interaction with the accretion environment during the frequent short
bursts \citep{Keek2016reflsim}. This opens up a new avenue to study
accretion processes. We saw that the already complex reflection models
will need to further take into account a wider range of compositions
and densities to take full advantage of such new X-ray burst observations
\citep{Ballantyne2004models,Garcia2016}.

\acknowledgements{L.K. is supported by NASA under award number NNG06EO90A. L.K. thanks
the International Space Science Institute in Bern, Switzerland for
hosting an International Team on X-ray bursts. This work benefited
from events supported by the National Science Foundation under Grant
No. PHY-1430152 (JINA Center for the Evolution of the Elements). This
research has made use of \emph{MAXI} data provided by RIKEN, JAXA,
and the \emph{MAXI} team. We thank the \emph{Swift} observatory for
performing the observations described in this paper. }

\bibliographystyle{apj}
\bibliography{sb1706}

\end{document}